\documentclass[aps,prd,eqsecnum,showpacs]{revtex4}

\usepackage{amssymb}
\usepackage{amsmath}
\usepackage{amsfonts}
\usepackage[dvips]{color,graphicx}
\usepackage{subfigure}

\begin{document}

\title{
Do Evaporating Black Holes Form Photospheres?}

\author{
$^{1}$Jane H. MacGibbon\footnote{Electronic address:  jmacgibb@unf.edu},
$^{2,3}$B.~J.~Carr\footnote{Electronic address:  B.J.Carr@qmul.ac.uk},
and $^{4}$Don N. Page\footnote{Electronic address:  don@phys.ualberta.ca} }

\affiliation{
$^{1}$Department of Chemistry and Physics, University of North Florida,
Jacksonville, Florida 32224, USA\\
$^{2}$Astronomy Unit, Queen Mary, University of London, Mile End Road, London
E1 4NS, UK\\
$^{3}$Research Center for the Early Universe, Graduate School of Science,
University of Tokyo, Tokyo 113-0033, Japan\\
$^{4}$Department of Physics, University of Alberta, Edmonton, Alberta T6G 2G7,
Canada }

\date{2008 Sept 15}


\begin{abstract}

Several authors, most notably Heckler, have claimed that the observable Hawking emission from a microscopic black hole is significantly modified by the formation of a photosphere around the black hole due to QED or QCD interactions between the emitted particles. In this paper we analyze these claims and identify a number of physical and geometrical effects which invalidate these scenarios. We point out two key problems. First, the interacting particles must be causally connected to interact, and this condition is satisfied by only a small fraction of the emitted particles close to the black hole. Second, a scattered particle requires a distance $\sim E/m_{e} ^{2} $ for completing each bremsstrahlung interaction, with the consequence that it is improbable for there to be more than one complete bremsstrahlung interaction per particle near the black hole. These two effects have not been included in previous analyses. We conclude that the emitted particles do not interact sufficiently to form a QED photosphere. Similar arguments apply in the QCD case and prevent a QCD photosphere (chromosphere) developing when the black hole temperature is much greater than $\Lambda _{QCD} $, the threshold for QCD particle emission. Additional QCD phenomenological arguments rule out the development of a chromosphere around black hole temperatures of order $\Lambda _{QCD} $. In all cases, the observational signatures of a cosmic or Galactic halo background of primordial black holes or an individual black hole remain essentially those of the standard Hawking model, with little change to the detection probability. We also consider the possibility, as proposed by Belyanin et al. and D. Cline et al., that plasma interactions between the emitted particles form a photosphere, and we conclude that this scenario too is not supported.

\end{abstract}

\pacs{04.70.Dy, 12.20.-m, 12.38.-t, 95.30.Cq, 98.70.-f
\hfill Alberta-Thy-XX-08}
 
Phys. Rev. {\bf D 78}, 064043 (2008)

\maketitle

\section{Introduction}

 In 1974 Hawking showed that black holes should continually emit radiation as a natural consequence of quantum field theory in curved spacetime \cite{H1,H2}. Hawking radiation is analogous to thermal radiation from a black body and leads to the identification of a classical black hole temperature. In the case of a Schwarzschild black hole, this temperature is inversely proportional to the black hole mass and therefore increases as the black hole radiates \cite{H1}. Only black holes with mass below $10^{26}{\rm \; g}$ have a temperature today greater than that of the cosmic microwave background. Such black holes may have been produced in the early Universe with dramatic cosmological consequences. Formation scenarios for primordial black holes (PBHs) include the collapse of overdense regions due to primordial inhomogeneities, especially those generated by inflation, a softening of the equation of state or bubble collisions at cosmological phase transitions, and the collapse of oscillating cosmic string loops or domain walls. The cosmological consequences include modifying the usual picture of cosmological nucleosynthesis, distorting the spectrum of the microwave background radiation, generating a cosmic baryon asymmetry, removing magnetic monopoles, and contributing to the dark matter. Even if PBHs never existed, studying them therefore places important constraints on models of the early Universe. All of these formation scenarios and cosmological consequences have recently been reviewed by Carr \cite{C05}.

 Of particular interest are those PBHs with initial mass $\sim 10^{15}{\rm \; g}$ which are evaporating at the present epoch, since these should contribute to the observable $\gamma$-ray, cosmic ray and neutrino backgrounds (\cite{CM98} and references therein). Indeed the detection of such backgrounds or of individual PBH bursts would be the first confirmation of the fundamental relationship linking black holes, general relativity and thermodynamics (see also Ref.~\cite{M1}). To date, no such PBH burst or background has been conclusively detected, although D. Cline et al. \cite{DC1,DC2,DC3} have attributed some short-timescale gamma-ray bursts to PBH explosions. However, observations of the 100 MeV gamma-ray background imply that $10^{15}{\rm \; g}$ PBHs must have $\Omega _{pbh} \lesssim 10^{-8} $ \cite{CM98}, where $\Omega _{pbh} $ is their present density in units of the critical density. The 5-year data from the Whipple gamma-ray telescope impose the strongest current limit on the PBH explosion rate \cite{Lin}. With the Standard Model of particle physics, this is much weaker than the gamma-ray background limit on $\Omega _{pbh} $ if the PBHs cluster smoothly within the Galactic halo, but it could be stronger if PBHs were highly clumped. The $\Omega _{pbh} $ limits from the Galactic cosmic rays are comparable with the gamma-ray background limit but depend on both the PBH clustering and the cosmic ray propagation model \cite{MC,Bar}.

 Black holes of $\sim 10^{15}{\rm \; g}$ have an initial temperature of about 20 MeV. Because a black hole will emit all particle species with rest mass less than or of order the black hole temperature, the emission from a 20 MeV black hole is sensitive to the threshold for quark and gluon production, $\Lambda _{QCD} \approx 200-300{\rm \; MeV}$. In the standard Hawking model \cite{MW}, the particles directly evaporated from a black hole are those which appear noncomposite compared to the wavelength of the radiated energy (or equivalently the black hole size) at a given temperature. These particles can then decay or form composite particles as the emission streams away from the black hole. A black hole with a temperature somewhat below $\Lambda _{QCD} $ should directly emit neutral and charged pions, as well as photons, electrons, positrons, muons, neutrinos and gravitons. Once the energy distribution of the Hawking emission significantly extends above $\Lambda_{QCD}$, the black hole should directly emit quarks and gluons rather than pions, with the quarks and gluons fragmenting and hadronizing in jets into the astrophysically stable species ($\gamma $, $p$, $\bar{p}$, $e^{+}$, $e^{-}$, $\nu $ and $\overline{\nu }$) after emission. This process is analogous to QCD jet decay in accelerators. By convolving the Hawking evaporation with the HERWIG Monte Carlo jet code, MacGibbon and Webber found that the particle flux from $0.3-100{\rm \; GeV}$ black holes is predominantly comprised of jet fragmentation products \cite{MW}. They found that, after decay, the flux peaks at an energy of around 100 MeV, roughly independent of the black hole temperature, with the average energy of the final species scaling more weakly than the black hole temperature. In contrast, prior to decay, the peak and average energies of the direct Hawking emission are proportional to the black hole temperature.

 In the Heckler photosphere scenario, as originally described in Refs.~\cite{HE1,HE2}, the situation is radically different. Above a certain black hole temperature, the particles are alleged to interact sufficiently after emission to form a quasithermal photosphere at a distance $\alpha ^{-4} $ times the Schwarzschild radius from the black hole \cite{HE1}, where $\alpha =e^{2} /\hbar c \approx 1/137$ is the fine structure constant. In this model, the relevant QED interactions are bremsstrahlung and electron-photon pair production. The critical temperature above which this happens is estimated to be $T_{crit} \sim 45$ GeV. When QCD effects are included, a lower transition temperature $T_{crit} \sim \Lambda _{QCD} /\alpha _{s} ^{5/2} $ is derived where $\alpha _{s} $ is the strong coupling constant: $\alpha _{s} \approx 0.12$ at energy scales around 100 GeV but it increases at lower energies and its precise behavior is unknown around $\Lambda _{QCD} $. Once the photosphere develops, the high energy emission is reprocessed to lower energies, drastically reducing the flux of high energy particles and significantly weakening the detectability of individual high-temperature black holes. Because the best limits on  $\Omega _{pbh} $ are derived from the present 100 MeV emission from a Galactic or extragalactic background of PBHs with temperatures well below $T_{crit} $, the limits on $\Omega _{pbh} $ are weakened only slightly by photosphere development \cite{HE2}. However, photosphere development would have considerable implications for experimental efforts to detect ultra-high-energy cosmic rays from the final explosive phases of PBHs in the present epoch \cite{Bug}. Indeed, this has led to decreased motivation among groups searching for such signatures.

 J. Cline, Mostoslavsky, and Servant \cite{CL} have numerically explored the development of the photosphere in the original Heckler model. They conclude that photosphere formation should set in above a slightly lower value of $T_{crit} $, weakening the bound on $\Omega _{pbh} $ still further and significantly changing the burst spectra. These authors, however, assume the same initial description of interactions as Heckler \cite{HE1,HE2}, so their work is not an independent derivation or confirmation of the model. Kapusta \cite{Kap1,Kap} and Daghigh and Kapusta \cite{DK1,DK2,DK3} have also performed calculations related to Heckler's model.

 In this paper we will raise a number of problems with the Heckler scenario. These include various geometrical effects, the form of the plasma mass correction, the validity of the perfect fluid assumption and QCD phenomenological considerations. However, the two most serious omissions in the Heckler scenario are as follows: (i) two particles must be in causal contact to interact, and this is satisfied only by a small fraction of the emitted relativistic particles when they are sufficiently close to the black hole that they would otherwise have significant interactions; and (ii) in a bremsstrahlung interaction the electron initially forms off-shell and must travel a distance in the black hole frame much greater than $\sim 1/m_e$  before completing each interaction and becoming on-shell. If it is `bumped' by encounters with other particles within $\sim 1/m_e$ of the hole, the Landau-Pomeranchuk-Migdal (LPM) effect implies that the particle generally undergoes at most one completed bremsstrahlung interaction as it streams away from the black hole. Thus we conclude that a persistent QED or QCD photosphere cannot form around a black hole. Rare interactions could slightly modify the $\gamma $, $p$, $\bar{p}$, $e^{+} $, $e^{-} $, $\nu $ and $\overline{\nu }$ signatures of PBHs but not significantly.

 Other authors, such Belyanin, Kocharovsky, and Kocharovsky \cite{BY} and D. Cline et al. \cite{DC1,DC3}, have proposed photosphere models which are unrelated to the Heckler mechanism. These models are mainly motivated by an attempt to explain certain short-period gamma-ray bursts. After analyzing the Belyanin-Kocharovsky-Kocharovsky and D. Cline et al. models, we conclude that particle interactions are also insufficient to form a QED or QCD photosphere in these scenarios. However, the relevance of PBH explosions at the present epoch should not be ruled out as some other PBH mechanism which has not yet been considered might explain the observed bursts.

 The plan of this paper is as follows. In Section II we review the fundamentals of Hawking evaporation. In Section III we outline the original Heckler photosphere scenario. In Section IV we analyze this scenario in detail. In Section V we summarize the results of our analysis and their observational implications. Other photosphere models are discussed in Section VI. Our conclusions are presented in Section VII. Except in Sections II and VIA, we will use Planck units in which $\hbar =k=G=c=4\pi \varepsilon _{0} =1$.

\section{HAWKING EVAPORATION}

 A Kerr-Newman black hole with angular velocity $\Omega $ and electric potential $\Phi $ radiates particles with total energy between $\left(E,E+dE\right)$ at a rate \cite{H1,H2}
\begin{equation} 
d\dot{N}_{s} =\sum _{n,l} \frac{\Gamma _{snl} dE}{2\pi \hbar } \left[\exp \left[\frac{E-n\hbar \Omega -e\Phi }{\hbar \kappa /2\pi c} \right]-\left(-1\right)^{2s} \right]^{-1}  
\label{1}
\end{equation} 
per degree of particle freedom. Here $s$ is the particle spin, $n\hbar $ is its axial quantum number or angular momentum, $l$ is its orbital angular momentum, $q$ is its charge and $\Gamma _{snl} $ is the dimensionless absorption probability for the emitted species. In general,  $\Gamma _{snl} $ is a function of  $E$, $\Omega $, $\Phi $ and the surface gravity of the black hole $\kappa $, as well as the internal degrees of freedom and rest mass of the emitted particle. Equation ~\eqref{1} implies that a charged, rotating black hole will preferentially emit particles which have the same sign of charge and spin as the black hole. The angular momentum of a black hole is emitted several times faster than its mass \cite{P2}. An electrically charged black hole quickly discharges provided $M_{bh} <10^{5} M_{\odot } $ , where $M_{bh} $ is the black hole mass \cite{Z,G,CA,P1,P3}. Thus we will henceforth assume an uncharged, nonrotating Schwarzschild black hole, i.e. $\Omega =\Phi =0$ and $\kappa =c^{4} / \left( 4GM_{bh} \right) $.

 The temperature of an uncharged, nonrotating black hole is \cite{H1}
\begin{equation}   
kT_{bh} =\frac{\hbar c^{3} }{8\pi GM_{bh} } =1.06\left(\frac{M_{bh} }{10^{13} {\rm g}} \right)^{-1} {\rm \; GeV}.  
\label{2}
\end{equation}
At all temperatures, a black hole will radiate the massless elementary particles: $s=1/2$ neutrinos (if we ignore neutrino mass), $s=1$ photons and $s=2$ gravitons. Massive elementary particles are evaporated in significant numbers once the peak in the energy distribution given by Eq.~\eqref{1} is of order their particle rest mass. Since the exponentially damped tail in Eq.~\eqref{1} extends to infinite energies, a nonzero contribution from any massive species is always present to some extent.

 At high energies, the sum over $n$ and $l$ of the absorption probabilities for both massless and massive species approaches the geometric optics limit \cite{P1,P2,P3}:
\begin{equation}
\Gamma _{s} \left(M_{bh} ,E\right)\equiv \sum _{n,l}\Gamma _{snl}  \approx \frac{27G^{2} M_{bh} ^{2} E^{2} }{\hbar ^{2} c^{6} } .
\label{3}
\end{equation}
At low energies, $\Gamma _{s} $  is suppressed for massless $s=0$ and $s=1/2$ species:
\begin{equation}
\Gamma _{s=0} \left(M_{bh} ,E\right)\approx \frac{16G^{2} M_{bh} ^{2} E^{2} }{\hbar ^{2} c^{6} } ,\qquad \Gamma _{s=1/2} \left(M_{bh} ,E\right)\approx \frac{2G^{2} M_{bh} ^{2} E^{2} }{\hbar ^{2} c^{6} } .
\label{4}
\end{equation}
For $s=1$ and $s=2$ massless bosons, the $E\to 0$ fall-off is steeper:
\begin{equation}
\Gamma _{s=1} \left(M_{bh} ,E\right)\approx \frac{64G^{4} M_{bh} ^{4} E^{4} }{3\hbar ^{4} c^{12} } ,\qquad \Gamma _{s=2} \left(M{}_{bh} ,E\right)\approx \frac{256G^{6} M_{bh} ^{6} E^{6} }{45\hbar ^{6} c^{18} } .
\label{5}
\end{equation}
The absorption probability for nonrelativistic massive particles is more strongly suppressed at low energies than for massless or relativistic particles. For $\left(M_{bh} /10^{13} {\rm g}\right)\left(\mu /{\rm GeV}\right)<10$, where $\mu $ is the particle rest mass energy, $\Gamma _{s=1/2} \left(M_{bh} ,E, \mu\right)$ at $E=\mu $ differs from the corresponding $\mu = 0$ value by up to 45\%  \cite{P3}. More importantly, since $E$ is the total energy, the spectrum of a massive species must be zero for $E<\mu $. These effects have a strong damping influence on the emission spectrum when the peak energy is close to the particle rest mass threshold.

 Prior to particle decays, the peaks in the flux per particle mode measured at infinity occur at \cite{P1,P2,P3,ES}
\begin{equation}  E_{s=0} \approx 2.81T_{bh} ,\qquad E_{s=1/2} =4.02T_{bh} , \qquad E_{s=1} =5.77T_{bh}  
\label{6}
\end{equation} 
and the total instantaneous flux emitted by a black hole is
\begin{equation} \frac{dN}{dt} =\sum _{i}{\rm n}_{i}  \int _{
\mu _{i}}^{\infty }\frac{d\dot{N}_{i} }{dE}  dE 
\label{7} 
\end{equation} 
where ${\rm n}_{i} $ is the number of degrees of freedom (or modes) per particle species $i$. The standard Glashow-Weinberg-Salam model with three generations has 2 modes per $s=1/2$ neutrino or antineutrino, 4 modes per $s=1/2$ $e^{\pm } $, $\mu ^{\pm } $ or $\tau ^{\pm } $ lepton, 12 modes per $s=1/2$ quark flavor, 16 modes per $s=1$ gluon, 2 modes per $s=1$  photon, and 2 modes  per $s=2$ graviton. With $40-100$ elementary modes for $T_{bh} \approx 0.3-100 {\rm \; GeV}$, the total emission rate of particles (prior to their decays) is
\begin{equation}
\frac{dN}{dt} \approx \frac{10^{-2} c^{3} }{GM_{bh} }.
\label{8}
\end{equation}

 Combining Eq.~\eqref{7} with the particle energy gives the instantaneous mass loss of the black hole. This can be integrated to derive the lifetime of a black hole of initial mass $M_{i} $ \cite{M2}:
\begin{equation}
\tau _{evap} =6.24 \times 10^{-27} M_{i} ^{3} f\left(M_{i} \right)^{-1} {\rm \; s}.
\label{9}
\end{equation}
The factor $f\left(M_{i} \right)$ depends on the number of emitted species and is normalized to unity when only massless species are emitted. Following the method of Ref.~\cite{M2}, we can now update the value of $M_{*} $, the initial mass of a black hole whose lifetime is the present age of the Universe. Using the 3-year Wilinson Microwave Anisotropy Probe (WMAP) best-fit value for the age of the Universe in the Lambda Cold Dark Matter (LCDM) model, $\tau _{u} =13.73_{-0.17}^{+0.13} \times 10^{9} {\rm \; yr}$ \cite{WMAP}, we find that
\begin{equation}
M_{*} =\left(5.00\pm 0.04\right)\times 10^{14} {\rm \; g}.
\label{10}
\end{equation}

 To aid our conceptual discussion of interactions between the emitted particles, we note the magnitudes of various physical quantities. The radius of the black hole is
\begin{equation}
r_{bh} =\frac{2GM_{bh} }{c^{2} } =1.49 \times 10^{-15} \left(\frac{M_{bh} }{10^{13} {\rm g}} \right){\rm \; cm}=1.57 \times 10^{-15} \left(\frac{T_{bh} }{{\rm GeV}} \right)^{-1} {\rm \; cm}.
\label{11}
\end{equation}
For directly emitted relativistic particles of energy $E$, the reduced de Broglie wavelength (a measure of the effective interaction range for interactions which fall off as some power of $1/E$) is \begin{equation}
\lambdabar _{b} =\frac{\hbar c}{E} =0.197 \times 10^{-13} \left(\frac{E}{{\rm GeV}} \right)^{-1} {\rm \; cm}.
\label{12}
\end{equation}
In the nonrelativistic case, the factor $E/c$ in Eq.~\eqref{12} is replaced by the momentum. Since the flux per $s=0,{\rm \; }1/2$ or 1 mode peaks at $E_{peak} \approx \left(3-6\right)T_{bh} $, we have
\begin{equation}
\lambdabar _{b} =\left(3-7\right) \times 10^{-15} \left(\frac{T_{bh} }{{\rm GeV}} \right)^{-1} {\rm \; cm}\approx \left(2-5\right)r_{bh}.
\label{13}
\end{equation}
A typical emitted particle therefore has a wavelength comparable to or somewhat larger than the size of the hole $2r_{bh} $. Additionally, the particles can be considered `evaporated' at a distance $r_{ev} \approx 3r_{bh} /2$ since the outgoing solution of the Hawking wave equation here already closely approximates the solution for the particle at infinity \cite{OH}.

 If the time between successive emissions from the black hole is less than $\lambdabar _{b} /c$, or equivalently if the distance travelled by a particle before the hole emits another particle is less than $\lambdabar _{b} $, we would expect interactions between the particles regardless of their species. Equation ~\eqref{8} gives the average time between successive emissions of elementary particles as
\begin{equation}
\Delta t\approx \frac{100GM_{bh} }{c^{3} },
\label{14}
\end{equation}
so
\begin{equation}
c\Delta t\approx \frac{20\hbar c}{E_{peak} } >>\lambdabar _{b}  
\label{15}
\end{equation} 
and such interactions are unlikely. More precisely, Oliensis and Hill have calculated numerically that more than 99\% of the particles emitted over a hole's lifetime satisfy the condition $\Delta t>\hbar /E$ \cite{OH}. This confirms the appropriateness of using $E_{peak} $ in Eq.~\eqref{15}.

 Similarly, from Eq.~\eqref{2}, once the temperature is above a few times $\Lambda _{QCD} $, the energy scale on which the QCD force becomes strong, the time between successive emissions is much shorter than the QCD timescale $\hbar /\Lambda _{QCD} $, and the size of the black hole $2r_{bh} $ is  much smaller than the associated distance $\hbar c/\Lambda _{QCD} $. Thus, for $T_{bh} >>\Lambda _{QCD} $, the Hawking evaporation of a colored particle is unaffected by the emission of other colored particles and the particle can be regarded as asymptotically free. On the scales associated with $\Lambda _{QCD} $ in the black hole rest frame, the emission of successive colored particles at these temperatures appears as effectively simultaneous production of highly energetic ($E>>\Lambda _{QCD} $) quarks and gluons at a point. This is analogous to QCD jet production in $e^{+}e^{-}$ annihilation accelerator events.  As the quarks and gluons propagate away from the black hole, they should fragment into further quarks and gluons and finally cluster into colorless hadrons once they have travelled a distance $\left(\hbar c/\Lambda _{QCD} \right)\left(E_{a} /\mu _{a} \right)$ from the black hole. Here $E_{a} $ represents an `average' energy of the jet particles involved in the decay and $\mu _{a} $ is an average `effective' rest mass energy.

 At temperatures above $\Lambda _{QCD} $, the observable spectra from the black hole, found by convolving the Hawking formula with jet decays, are dominated by the QCD decay products. Once the emitted particles have decayed into the astrophysical stable species, the average energy for $T_{bh} \approx 0.3-100{\rm \; GeV}$  scales as approximately $T_{bh} ^{1/2} $ for $\gamma $, $e^{+} $, $e^{-} $, $\nu $ and $\overline{\nu }$, and $T_{bh} ^{0.8} $ for $p$ and $\overline{p}$, rather than as $T_{bh} $ as given by Eq.~\eqref{6} \cite{MW}. The particle flux after decay, $dN/dt$, for $T_{bh} \approx 0.3-100{\rm \; GeV}$ scales as approximately  $T_{bh} ^{1.6} $ \cite{MW}, rather than as $T_{bh} $ as implied by Eq.~\eqref{8}.

\section{THE HECKLER SCENARIO}

 We first focus on the QED interactions in the Heckler scenario, as originally derived in Refs.~\cite{HE1, HE2}. Consider a $T_{bh} >>m_{e} $ black hole which is emitting Hawking radiation in random radial directions. From Eq.~\eqref{1} the number density at radius $r$ of electrons and positrons which were directly radiated by the black hole (and not the result of particle decays) is
\begin{equation}
n_{0} \left(r\right)\approx \frac{10^{-4} }{M_{bh} r^{2} } ,
\label{16}
\end{equation}
where we are now using units with $\hbar =k=G=c=1$. If the number density is sufficiently high, the electromagnetic particles could lose energy via two-body bremsstrahlung $e+e\to e+e+\gamma $ and pair production $e+\gamma \to e+e^{+} +e^{-} $ as they propagate away from the hole. Each such interaction would increase the number of electromagnetic particles by a factor of $3/2$.

 To calculate the number of scatterings per particle, Ref.~\cite{HE1} takes the bremsstrahlung cross-section in the center-of-mass frame --- averaged over the photon energy $\omega $ --- to be \cite{JR,EH1}
\begin{equation}
\sigma _{brem} =\frac{1}{E} \int _{0}^{E}\omega \frac{d\sigma _{brem} }{d\omega } d\omega  \approx \frac{8\alpha ^{3} }{m_{e} ^{2} } {\rm ln}\frac{2E}{m_{e} } .
\label{17}
\end{equation}
The center-of-mass frame for each interaction is assumed to be the rest frame of the black hole. In Eq.~\eqref{17}, $E\approx 4T_{bh} $ is the energy of the initial electron or positron. Since the average momentum exchanged by the virtual photon in the bremsstrahlung interaction is $\sim m_{e} $, the Heisenberg Uncertainty Principle implies that the two initial particles must be within $\sim 1/m_{e} $ of each other to interact significantly. In Ref.~\cite{HE1} the `formation' distance for the final on-shell electron and photon is also assumed to be $\sim 1/m_{e} $ in the center-of-mass frame of the interaction. In the relativistic limit, the bremsstrahlung cross-sections for two incident electrons or two incident positrons or an electron and a positron are the same. The pair-production cross-section can be derived from the bremsstrahlung cross-section using the substitution law (see Section IV.K) and has the same functional form in the relativistic limit \cite{EH2}.

  In Ref.~\cite{HE1}, $m_{e} $ in Eq.~\eqref{17} is replaced by an effective electron mass $m'_{e} $ which is taken to be the vacuum electron mass $m_{e} $ augmented by a `plasma mass' $m_{pm} $ to account for Coulomb screening by the background of other nearby emitted electromagnetic particles. The electron mass is approximated by the correction applicable for an isotropic thermal plasma
\begin{equation} 
m'_{e} =\sqrt{m_{e} ^{2} +m_{pm} ^{2} } ,\qquad  m_{pm} ^{2} \approx \frac{4\pi \alpha n\left(r\right)}{E_{av} },
\label{18}
\end{equation} 
which includes the further approximations $\left(1/E\right)_{av} \approx 1/E_{av} \approx 1/4T_{bh} $. This increases the effective electron mass, resulting in a decreased cross-section, within a distance $\sim 0.1\alpha ^{1/2} m_{e} ^{-1} $ of the black hole. However, there is an inaccuracy here in that the emitted particles do not form an isotropic thermal plasma, as pointed out by Heckler \cite{HE1} and Cline, Mostoslavsky, and Servant \cite{CL} themselves.

 The number of scatterings, ${\mathcal N}$, experienced by a particle by the time it reaches radius $R$ is found by calculating
\begin{equation}
{\mathcal N}\left(R\right)=\int _{r_{\min } =r_{bh} }^{r_{\max } =R}\frac{dr}{\lambda \left(r\right)}
\label{19}
\end{equation}
with $\lambda \left(r\right)$, the mean-free-path of an individual particle, defined as 
\begin{equation}
\lambda \left(r\right)=\left(n\left(r\right)\sigma _{brem} v_{rel} \right)^{-1} 
\label{20}
\end{equation}
and
\begin{equation}
n\left(r\right)=\left(\frac{3}{2} \right)^{{\mathcal N}\left(r\right)} n_{0} \left(r\right).
\label{21}
\end{equation}
The factor of $3/2$ accounts for the increase in the number of electromagnetic particles at each bremsstrahlung interaction, and the relative velocity between the interacting particles is taken to be $c$. Only the lower limit in integral \eqref{19} is important. As $r\to \infty $, the number density of the emission decreases as $r^{-2} $ and the scattering cuts off.  Heckler first neglects the plasma mass correction and takes the lower limit in integral \eqref{19} to be $1/m_{e} $. In this case, the integral becomes
\begin{equation}
{\mathcal N}\approx \frac{\alpha ^{3} }{2\pi ^{4} } \frac{T_{bh} }{m_{e} } \ln \frac{2T_{bh} }{m_{e} } ,
\label{22}
\end{equation}
which exceeds $1$ for
\begin{equation}
T_{bh} \gtrsim \frac{\pi ^{2} }{\alpha ^{3} } m_{e} \approx 20{\rm \; TeV}.    
\label{23}
\end{equation}
The analysis is then extended to include the expression for the plasma mass given by Eq.~\eqref{18} and the lower limit in integral \eqref{19} is taken to be the radius of the black hole. For $r$ small enough that $m_{pm} \left(r\right)>m_{e} $, the number of scatterings is damped by the plasma mass. The dominant contribution to the integral then comes from the value of $r$ at which $m_{pm} \approx m_{e} $ and this gives 
\begin{equation}  
{\mathcal N}\approx \frac{\alpha ^{5/2} }{\pi ^{3/2} } \frac{T_{bh} }{m_{e} } \ln \frac{T_{bh} }{m_{e} } ,
\label{24}
\end{equation} 
which exceeds 1 for 
\begin{equation}
T_{bh} \gtrsim \frac{\pi ^{3/2} }{\ln (\alpha ^{-5/2} )} \frac{m_{e} }{\alpha ^{5/2} } \approx 45{\rm \; GeV}.
\label{25}
\end{equation}
On numerically integrating Eq.~\eqref{19}, Heckler finds that ${\mathcal N}=1$ when $T_{crit} \approx 45.2{\rm \; GeV}$. Thus, although the cross-section $\sigma _{brem} \propto \alpha ^{3} $ remains small, the growth in number density around the black hole at high $T_{bh} $ leads to significant interaction in this model.

 The region for which ${\mathcal N}>>1$ is the photosphere region. It is claimed that when $T_{bh} >>T_{crit} $ the multiple scatterings are sufficient to produce an interacting quasithermal fluid, which flows away from the black hole. The photosphere is found to have inner and outer radii given by 
\begin{equation}
r_{in} \approx \frac{4\pi }{\alpha ^{4} T_{bh} } =\frac{16\pi ^{2} }{\alpha ^{4} } r_{bh} \approx 10^{9} r_{bh} ,\qquad
r_{out} \approx \frac{1}{\alpha ^{2} m_{e} } \left(\frac{T_{bh} }{T_{crit} } \right)^{1/2} .
\label{26}
\end{equation}
Approximating this region by a perfect fluid, the average energy of the particles at the outer boundary of the photosphere, in effect the observable average energy of the emitted particles, is estimated to be
\begin{equation}
E_{obs} \approx m_{e} \left(T_{bh} /T_{crit} \right)^{1/2} ,
\label{27}
\end{equation}
which is considerably less than the energy $E_{av} \approx 4T_{bh} $ expected without photosphere development. (These values do not include the electrons and photons produced by the decays of other particles radiated by the black hole.) Therefore the flux of photons observed from a black hole with a QED photosphere is 
\begin{equation}
\frac{dN_{\gamma } }{dt} \approx \frac{T_{bh} }{\alpha ^{2} } \left(\frac{T_{bh} }{T_{crit} } \right)^{1/2},
\label{28}
\end{equation}
neglecting particle decays, compared to  $dN_{\gamma } /dt\approx 10^{-2} T_{bh} $ without a photosphere.

 Extending the Heckler model to QCD effects, Eq.~\eqref{17} is replaced in Ref.~\cite{HE1} by the cross-section for the gluon-bremsstrahlung process $q+q\to q+q+g$:
\begin{equation}  
\sigma _{brem} \approx \frac{8\alpha _{s} ^{3} }{m_{q} ^{2} } {\rm ln}\frac{2E}{m_{q} }
\label{29}
\end{equation} 
where $\alpha _{s} $ is now the strong coupling constant. The effective quark mass, $m_{q} $, is also enhanced by a plasma term:
\begin{equation}
m_{q} ^{2} =\Lambda _{QCD} ^{2} +\frac{\alpha _{s} \left({\mathcal N}_{ QCD} +1\right)^{2} }{\left(4\pi ^{2} r\right)^{2} }.
\label{30}
\end{equation}
Proceeding as before, one arrives at a rough estimate for the onset of a photosphere due to QCD interactions: ${\mathcal N}_{QCD} >>1$ when $T_{bh} $ exceeds
\begin{equation}
T_{crit}^{QCD} \approx \Lambda _{QCD} /\alpha _{s} ^{5/2} \gtrsim \Lambda _{QCD} .
\label{31}
\end{equation}
The precise value of $T_{crit}^{QCD} $ is sensitive to threshold effects and the energy dependence of $\alpha _{s} $ around $\Lambda _{QCD} $. Because $T_{bh} $ must be greater than about $\Lambda _{QCD} $ for a black hole to emit QCD particles, $T_{crit}^{QCD} $ cannot be less than about $\Lambda _{QCD} $.

Cline, Mostoslavsky, and Servant \cite{CL} adapted a code developed to study heavy ion collisions to obtain numerical solutions of the Boltzmann equation in the context of Heckler's model.  As discussed in Section IV.L, these authors found that QED and QCD photospheres may form but disagreed with some features of the original Heckler model. Once the photosphere formed, they also considered the effect of elastic Compton scattering (which conserves particle number) and found that this can extend the photosphere when $T_{bh} \gtrsim 5 {\rm \; TeV}$. However, because Cline, Mostoslavsky, and Servant employed the same initial description of interactions as Heckler, their work is not an independent derivation or confirmation of photosphere development.

\section{IS THE HECKLER SCENARIO CORRECT?}

In this section, we show that several aspects of the original Heckler model are invalid and conclude that QED photosphere development is ruled out. We begin by focusing on the QED bremsstrahlung interactions. Extension to QED pair production, as well as the corresponding QCD interactions, is straightforward and discussed in Section IV.K. As in the original Heckler model, we will consider all emitted particles to be initially moving in random radial directions away from the black hole. The effect of black hole size on the emission directions is negligible for $r>>r_{bh} $. We will also ignore general relativistic effects, which are negligible for $r \gtrsim 10r_{bh} $.

\subsection{$\sigma _{brem} $ in More Detail}

\begin{figure}[t]
\begin{center}
\rotatebox{-90}{
\includegraphics[width=0.60\linewidth]{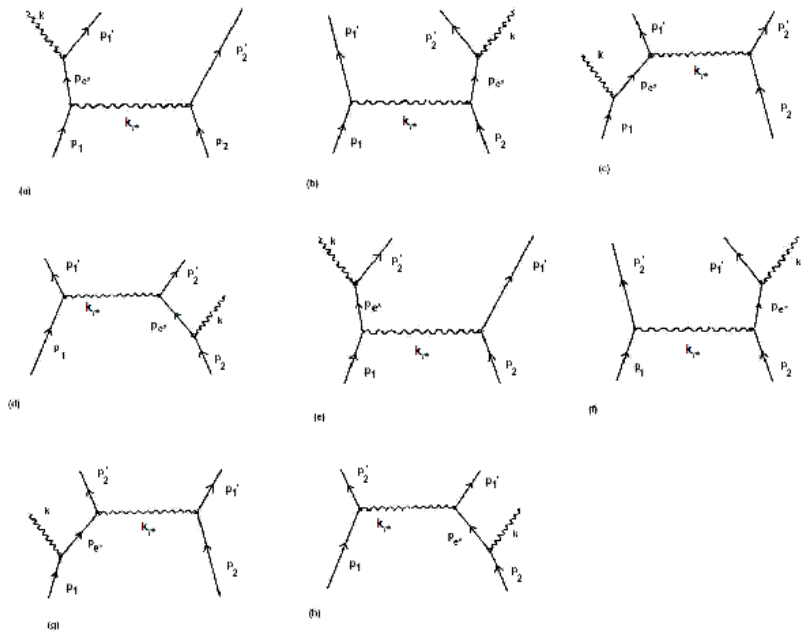}
}
\caption{\label{F1}
The eight Feynman diagrams for electron-electron bremsstrahlung in lowest order perturbation theory. The labels, which are defined in the text, correspond to the 4-momenta of the particles involved in the interaction.
}
\end{center}
\end{figure}

 First we consider whether the bremsstrahlung cross-section used by Heckler is appropriate. The double-differential cross-section for two-body electron bremsstrahlung is given in Ref.~\cite{EH1}. In this interaction, two electrons and/or positrons exchange a virtual photon, forcing one electron off-shell. The off-shell electron then decays into an on-shell electron by emitting a photon. The eight relevant Feynman diagrams for lowest order perturbation theory are shown in Fig. ~\ref{F1} \cite{AND}. The cross-section Eq.~\eqref{17} applies in the relativistic limit for $e^{-} e^{-} $, $e^{+} e^{+} $ and $e^{-} e^{+} $ bremsstrahlung because the interactions associated with the exchange diagrams, Figs. 1 (e), (f), (g) and (h), are suppressed at high energy. (In the nonrelativistic limit, the $e^{-} e^{+} $ bremsstrahlung cross-section is greater than the $e^{-} e^{-} $ and $e^{+} e^{+} $ cross-sections due to the dipole nature of $e^{-} e^{+} $ \cite{JR}.) In the relativistic limit, the average angle between the final on-shell electron and photon is \cite{EH1}
\begin{equation} 
\left|\phi \right|_{av} \approx \frac{m_{e} }{2E}  
\label{32}
\end{equation} 
in the center-of-mass frame of the interaction, where $E$ is the energy of each initial electron. The average final energies of the on-shell electron and photon are \cite{EH1}
\begin{equation}
E_{e} \approx \omega \approx E/2.
\label{33}
\end{equation}
Because $\phi _{av} $ is so small, the majority of interactions involve very small momentum transfer, typically of order $m_{e} $. Thus the directions of the electrons are modified little by the interaction but their energy loss can be substantial.

 Since the average energy of the final photon is of order $E$, it is valid to use the energy-averaged two-body bremsstrahlung cross-section given by Eq.~\eqref{17} which subsumes the infrared divergence as $\omega \to 0$. In Ref.~\cite{EH1}, $d\sigma _{brem} /d\omega $ is evaluated by numerically integrating the complicated double-differential cross-section. The cross-section does not contain the collinear divergences which would arise at $\phi =0{}^\circ $  and $\phi =180{}^\circ $ if factors of order $m_{e} $ were not correctly included in the derivation. In the non-relativistic limit, $\sigma _{brem} $ decreases more strongly with decreasing $E$ than Eq.~\eqref{17} \cite{EH1} and becomes process-dependent \cite{JR,EH3}, as noted above.

It should be stressed that the Heckler effect is not due to an infrared or collinear divergence in the cross-section but to the high number density of emitted particles flowing isotropically and relativistically away from the black hole. As a consequence of the emission distribution, the center-of-mass frames for most of the interactions are not moving highly relativistically with respect to the black hole, as we next show.

\subsection{Non-Interaction of Radially Comoving Particles}

 In Section II, we saw that one should not expect QED or QCD interactions between emitted particles if the interaction scale falls off as $1/E$ (i.e. if the cross-section falls off as $1/E^{2} $). In the Heckler scenario a quasithermal photosphere develops from QED and QCD bremsstrahlung and pair-production interactions, which do not obey this constraint. We now demonstrate that particles moving in similar radial directions do not interact significantly even via bremsstrahlung or pair production.

\begin{figure}[t]
\begin{center}
\rotatebox{90}{
\includegraphics[width=0.30\linewidth]{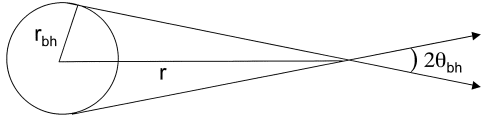}
}
\caption{\label{F2}
Illustrating how a particle emitted by a black hole of radius $r_{bh} $ can interact with other particles emitted in a similar radial direction where the angle $\theta _{bh} $ is the deviation from the radial direction due to the finite size of the black hole.
}
\end{center}
\end{figure}

 Consider a typical directly emitted particle of rest mass $\mu <<T_{bh} $ and energy $E\approx 4T_{bh} $ at a distance $r>>r_{bh} $ from the black hole. At this distance, special but not general relativistic effects are relevant. In the rest frame of the black hole, the particle has an individual Lorentz factor
\begin{equation} 
\gamma \approx \left(2\pi M_{bh} \mu \right)^{-1}
\label{34}
\end{equation} 
and velocity
\begin{equation}
v\approx 1-2\pi ^{2} M_{bh} ^{2} \mu ^{2} .
\label{35}
\end{equation}
To maximize the possible effect, we allow for the small deviations from the radial direction due to the finite size of the black hole. From simple geometry (see Fig. ~\ref{F2}), the particle will be moving at an average angle
\begin{equation} 
\theta _{bh} \left(r\right)\approx \frac{r_{bh} }{r}
\label{36} 
\end{equation} 
from the radial direction. Note that, for $r>>r_{bh} $, almost all of the velocity is radial and $\theta _{bh} \to 0$ as $r\to \infty $. The transverse component of the particle velocity in the black hole rest frame is
\begin{equation}
v_{T} \approx \frac{r_{bh} }{r} ,
\label{37}
\end{equation}
giving a radial velocity component of
\begin{equation}
v_{R} \approx 1-2\pi ^{2} M_{bh} ^{2} \mu ^{2} -\frac{2M_{bh} ^{2} }{r^{2} } .    
\label{38}
\end{equation}
The radial velocity is approximately equal to $v_{CM} $, the velocity of the center-of-mass (CM) of two typical particles moving at an angle $\theta _{bh} $ either side of the radial direction. The corresponding Lorentz factor for the CM of the particles is 
\begin{equation}
\gamma _{CM} \approx \frac{r}{2M_{bh} } \left(1+\pi ^{2} \mu ^{2} r^{2} \right)^{-1/2}.    \label{39}
\end{equation}
At large $r$, this reduces to $\gamma _{CM} \approx 4T_{bh} /\mu $, as expected. The special relativistic transformation of velocities is
\begin{equation}
v_{x} '=\frac{v_{x} -V}{1-Vv_{x} }, \qquad
v_{y} '=\frac{v_{y} \sqrt{1-V^{2} } }{1-Vv_{x} } ,
\label{40}
\end{equation}
where $V$ is the relative velocity of the reference frames, taken to be in the $x$ direction. Thus the relative velocity of the two particles in their CM frame is 
\begin{equation}
v'_{rel} \approx \left(1+\pi ^{2} \mu ^{2} r^{2} \right)^{-1/2} ,
\label{41}
\end{equation}
where the prime indicates the CM frame. This is close to $1$ for radii less than $\mu ^{-1} $ but very small at distances much larger than this.

 The total number density of emitted particles at a radius $r$ in the black hole rest frame is given by Eq.~\eqref{16}. In the CM frame of the two particles, the number density at  $r$ is therefore
\begin{equation}
n'\left(r\right)\approx \frac{10^{-3} }{\gamma _{CM} M_{bh} r^{2} } \approx 10^{-3} r^{-3} \left(1+\pi ^{2} \mu ^{2} r^{2} \right)^{1/2} .
\label{42}
\end{equation}
Hence the number of interactions per particle, integrated over all radii, is
\begin{equation}
{\mathcal N}\approx \int n'\sigma v'_{rel}  dt'=\int \frac{n'\sigma v'_{rel} }{\gamma _{CM} }  dr.  \label{43}
\end{equation}
Taking the cross-section in the CM frame to be the electron bremsstrahlung cross-section (the pair-production cross-section is similar), and allowing for the plasma mass correction (see Section III), we find
\begin{equation}
\sigma _{brem} \approx \frac{10\alpha ^{3} }{m_{e} ^{2} +\frac{\alpha }{10^{3} r^{2} } } ,     \label{44}
\end{equation}
which gives
\begin{equation}
{\mathcal N}\approx \int _{r_{bh} }^{\infty }10^{-2} r^{-3}  \left(1+\pi ^{2} m_{e} ^{2} r^{2} \right)^{1/2} \alpha ^{3} \left(m_{e} ^{2} +\frac{\alpha }{10^{3} r^{2} } \right)^{-1} M_{bh} r^{-1} dr 
\label{45}
\end{equation} 
for $T_{bh} >> m_{e} $. The dominant contribution to the integral comes from the lower limit, here taken to be the radius of the black hole. Near this limit, the second term in the first brackets and the first term in the second brackets can be neglected, so we obtain
\begin{equation}
{\mathcal N}\approx \int _{r_{bh} }^{\infty }10 \alpha ^{2} M_{bh} r^{-2} dr\approx 10\alpha ^{2} <<1 .
\label{46}
\end{equation}
Hence emitted particles traveling in a similar radial direction will not interact with each other via bremsstrahlung or pair production, or scatter or lose energy to any significant degree, as they propagate away from the black hole. This is a consequence of the Lorentz factor of the emitted particles being so large in the black hole frame.

\subsection{Center-of-Momentum Frame of Interactions}

 We now examine the Heckler assumption that the CM frame for most interactions is the rest frame of the black hole, so that $\sigma _{brem} $ is given by Eq.~\eqref{17} in the black hole frame. As we have seen, this is not true for particles moving in similar radial directions. Here we investigate particles moving in different radial directions.

 Consider ultra-relativistic particles 1 and 2 of the same energy and rest mass, created near the black hole at the same time and subsequently moving radially outwards with speed $v_1$ along rays separated by an angle $\theta $. The Lorentz factor $\gamma _{CM} $ of the CM of the two particles relative to the rest frame of the black hole is
\begin{equation}
\gamma _{CM} =\left(1-v_{1} ^{2} {\rm cos}^{{\rm 2}} \frac{\theta }{2} \right)^{-1/2} \approx \left({\rm sin}\frac{\theta }{2} \right)^{-1} .
\label{47}
\end{equation} 
This is approximately independent of $v_{1} $ (and hence $T_{bh} $) when $\theta \gtrsim 2/\gamma _{1} $. Equation ~\eqref{47} implies that $\gamma _{CM} =100$ when $\theta \approx 1{}^\circ $; $\gamma _{CM} =10$ when $\theta \approx 11.5{}^\circ $; $\gamma _{CM} \approx 3$ when $\theta =39{}^\circ $; and $\gamma _{CM} =1$ when $\theta =180{}^\circ $. Thus the CM frames for the interactions of particle 1 with the majority of particles streaming out from the black hole are not significantly Lorentz-boosted relative to the black hole. The region for which the interaction frames are significantly Lorentz-boosted relative to the black hole is a cone of angle $<<\pi /2$ around the direction of particle 1.

 In general, the energies of particles 1 and 2 are not equal but have the Hawking distribution, given by Eq.~\eqref{1}, and the corresponding formula for $\gamma _{CM} $ is more complicated. Let us denote the ultra-relativistic speeds of particles 1 and 2 by $v_{1} =1-\varepsilon _{1} $ and $v_{2} =1-\varepsilon _{2} $, respectively, where $\varepsilon _{1} ,\varepsilon _{2} <<1$, and let $\theta _{CM1} $ be the angle between $\boldsymbol{v}_{1} $ and the velocity of the CM of particles 1 and 2, $\boldsymbol{v}_{CM} $. The special relativistic equations for relative velocity then give 
\begin{equation}
v_{CM} \approx \frac{1-\left(\varepsilon _{1} +\varepsilon _{2} \right)}{\left(\left(1-{\rm cos\; }\theta \right)\left(\frac{{\rm sin}\theta _{{\rm CM1}} }{{\rm sin}\theta } \right)+{\rm cos}\theta _{CM1} \right)-\left(\varepsilon _{1} -\varepsilon _{2} {\rm cos}\theta \right)\left(\frac{{\rm sin}\theta _{{\rm CM1}} }{{\rm sin}\theta } \right)-\varepsilon _{2} {\rm cos}\theta _{CM1} }
\end{equation}
and
\begin{equation}
\frac{{\rm sin}\theta _{{\rm CM1}} }{{\rm sin}\theta } \approx \left(\frac{\varepsilon _{1} \left(\left(\varepsilon _{1} +\varepsilon _{2} \right)-2\sqrt{\varepsilon _{1} \varepsilon _{2} } {\rm cos}\theta \right)}{\left(\varepsilon _{1} -\varepsilon _{2} \right)^{2} +4\varepsilon _{1} \varepsilon _{2} {\rm sin}^{2} \theta } \right)^{1/2} 
\label{48}
\end{equation}
for $0{}^\circ \le \theta \le 180{}^\circ $. Again $\gamma _{CM} =\left(1-v_{CM} ^{2} \right)^{-1/2} $ is approximately independent of $v_{1} $ and $v_{2} $ (and thus $T_{bh} $) unless $\theta $ is extremely close to zero. If we assume $E_{1} \le E_{2} $, without loss of generality, and consider $E_{2} \le 2E_{1} $, corresponding to $\varepsilon _{2} \le \varepsilon _{1} /4$ (this more than adequately covers the majority of the Hawking distribution around $E_{peak} $), we have $\gamma _{CM} \le 3$ when $\theta \gtrsim 42{}^\circ $ and $\gamma _{CM} \le 10$ when $\theta \gtrsim 12{}^\circ $. Similarly if $E_{2} \le 4E_{1} $, then $\gamma _{CM} \le 3$ when $\theta \gtrsim 50{}^\circ $ and $\gamma _{CM} \le 10$ when $\theta \gtrsim 15{}^\circ $.

Hence we conclude that taking the CM frame of each interaction to be the black hole rest frame, as Heckler does, is a valid approximation for the interactions of any given particle with most other particles. To within a factor of $O\left(1\right)$, using $\sigma _{brem} $ in the black hole frame to describe the interactions of the particle is correct, except for an exclusion cone of angle $\theta _{ex} \approx O\left(0.1-1\right)$ radians around the direction of the particle's velocity for which the interactions are negligible because $\gamma _{CM} \left(\theta \right)>>1$. However, as we shall see below, this exclusion cone has important consequences which hinder the development of the photosphere.

\subsection{Transverse Distance}

 Once the particle has traveled a distance $d$ from the black hole, the transverse distance between the particle and the edge of its exclusion cone is
\begin{equation}
x_{T} \approx d\theta _{ex} \approx O\left(d\right) .
\label{49}
\end{equation}
Because $x_{T} $ is a transverse distance, $x_{T} $ is the same in the particle and black hole frames. Thus once the particle has traveled a distance $r_{brem} \approx 1/m_{e} $ from the black hole, the distance to the nearest particle with which it could then interact is of order $1/m_{e} $, provided the emitted particles are still traveling mainly radially. Because $\sigma _{brem} $ dictates that two particles must be within $\sim 1/m_{e} $ of each other to interact via bremsstrahlung, few interactions can occur after the particles have propagated this distance. The relevant interactions for initiating photosphere formation can only be those which occur within $r_{brem} $ of the black hole. This point is not included in the original Heckler model.

As we shall demonstrate in Section IV.G, the electron coming out of a bremsstrahlung interaction is still off-shell at $r_{brem} $. Also $\sigma _{brem} $ is truncated by the causality constraint and may be damped by off-shell (LPM-type) interactions, as we discuss in Sections IV.F and IV.H. These effects significantly decrease the capacity for photosphere formation.

\subsection{Deviations from the Radial Direction}

 We have assumed in the exclusion cone argument that the emitted particles move out radially from the black hole. We now justify this. Were they to random walk out, an individual particle would have to undergo
\begin{equation} 
{\mathcal N}_{ex} \approx \left(\frac{\theta _{ex} }{\phi _{av} } \right)^{2}  
\label{50}
\end{equation} 
scatterings to deviate by an angle $\theta _{ex} \approx O\left(0.1-1\right)$ from the radial direction, where $\phi _{av} $ is given by Eq.~\eqref{32}. If $T_{bh} \approx 1-10$ GeV, ${\mathcal N}_{ex} \gtrsim 10^{7} $ scatterings are required for the particle to scatter outside the exclusion cone. Thus even if a photosphere can develop, particles will deviate little from the radial direction before and after photosphere formation. (As discussed in Section IV.L, this is incorrectly treated in the numerical work of Cline, Mostoslavsky, and Servant \cite{CL}.) Therefore the constraint of Section IV.D set by the transverse distance to the edge of the exclusion cone will also apply after the onset of scattering. This confirms that the only relevant interactions can be those which occur within $r_{brem} $ of the hole.

Additionally, if we consider the possible deviation from the radial direction due to the finite size of the black hole, Eq.~\eqref{36} implies that $\theta _{bh} \approx \theta _{ex} $ only when $r\approx r_{bh} $. This is approximately the radius at which the Hawking emitted particle first appears and, for $T_{bh} >>m_{e} $, represents a tiny portion of the region within $r_{brem} $ of the black hole. Hence the effect of black hole size on the propagation direction should also be negligible.

\subsection{Geometrical and Causality Considerations}

\begin{figure}[t]
\begin{center}
\rotatebox{-90}{
\includegraphics[width=0.30\linewidth]{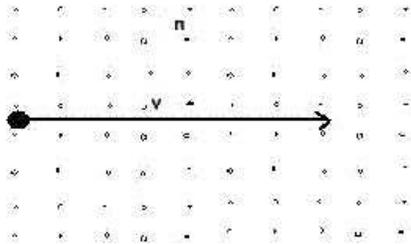}
}
\caption{\label{F3}
The distribution of background particles for which the formula $\lambda =\left(n\sigma v\right)^{-1} $ describes the mean-free-path of an incident particle of speed $v$. The number density $n$ is assumed to be uniform. In the actual black hole situation the density is not uniform and all particles are moving radially outwards in the black hole frame.}
\end{center}
\end{figure}

 We now calculate the number of interactions ${\mathcal N}$ each particle undergoes. The definition of the mean-free-path used in the Heckler model and given by Eq.~\eqref{20} is clearly wrong for geometrical reasons. Equation ~\eqref{20} applies when a particle propagates linearly through a target of uniform number density $n$ in the configuration shown in Fig. ~\ref{F3} \cite{MK}. This is not geometrically analogous to the black hole situation for three reasons. First, the particles are moving radially outward in a spherical distribution from the black hole. Second, the number density $n\left(r\right)$ varies with the distance of the moving particles from the black hole. That is, if we consider particle 1 interacting with an element of the flux around particle 2, $n\left(r_{2} \right)$ does not remain constant as particles 1 and 2 travel outwards. Third, implicit in Eq.~\eqref{20} is the assumption that the interacting particles have an infinite past and future, which is not the case for black hole emission. Consider particle 1 at radial position $\boldsymbol{r}_{1} $ and particle 2 at $\boldsymbol{r}_{2} $ with $|\boldsymbol{r}_{2} |<|\boldsymbol{r}_{1} |$. Particle 2 was emitted by the black hole at a time  $\Delta t=|\boldsymbol{r}_{2} |-|\boldsymbol{r}_{1} |$ later than particle 1. The Hawking derivation of Eq.~\eqref{1} utilizes the solutions at infinity but does not tell us how a particle is evaporated: either the particle did not exist or it was hidden by the event horizon before emission. In either interpretation, the particle prior to its emission cannot interact with particles outside the event horizon. The interaction lifetime of the particle is therefore truncated. Because ${\mathcal N}$ in Eqs.~\eqref{20} and \eqref{24} is dominated by the interactions of particle 1 with the short-lived, high density particles close to the black hole, this truncated lifetime should have a significant effect on the Heckler scenario.

An important consequence of the truncated lifetime is its implication for causality. From the classical point of view, i.e. ignoring temporarily the quantum smearing of particle space-time position and energy-momentum due to the Heisenberg Uncertainty Principle and the quantum nature of interactions, two particles can interact only if they are causally connected or, equivalently, if a light signal can be transmitted between them.  Classically each particle is point-like and emitted by the black hole at a precise time. Without loss of generality, let us assume particle 1, emitted at time $t_{1} $ in the black hole rest frame, is the scattering particle in the bremsstrahlung interaction and particle 2, emitted at $t_{2} =t_{1} +\Delta t$, is the scattered particle. The time difference $\Delta t$ may be positive or negative. The most appropriate frame for analyzing the causality constraint is the frame of the scattering particle because the Coulomb field of the scattering particle has an undistorted $e/r^{2} $ distribution in its own frame. Let us consider a light signal which travels a distance $D$ from the scattering particle in the scattering particle's frame and reaches the scattered particle. The form of the invariant $\sigma _{brem} $, Eq.~\eqref{17}, implies that, if the particle histories were not truncated, the probability for the scattered particle to undergo a bremsstrahlung event would be $\sim \alpha ^{3} $ if $D\lesssim 1/m_{e} $, but negligible if $D>>1/m_{e} $. (The truncation of the particle histories will further suppress the probability within $D\lesssim 1/m_{e} $ \cite{PCM}.) If $\Delta t>0$ and particle 1 emits a light signal which reaches particle 2 just as particle 2 is emitted by the black hole, one can readily calculate \cite{PCM} that
\begin{equation} 
D=\sqrt{\gamma ^{2} -1} {\rm \; }\Delta t=\gamma v\Delta t 
\label{51} 
\end{equation} 
where $v$ is the velocity of the particles in the black hole frame. If, alternatively, $\Delta t<0$ and particle 1, as soon as it is evaporated by the black hole, emits a light signal which then reaches particle 2, a slightly longer calculation \cite{PCM} shows that
\begin{equation} 
D=\gamma v\left[\frac{1-v\cos \theta }{1-v} \right]{\rm \; }\left|\Delta t\right|>\gamma v\left|\Delta t\right| 
\label{52} 
\end{equation} 
where $\theta $ is the angle between the particles in the black hole frame. Thus in either case, for the $D \lesssim 1/m_{e} $ constraint to be met, the difference in the emission times must be
\begin{equation} 
\Delta t \lesssim \Delta t_{c} =\frac{1}{\gamma vm_{e} } \approx \frac{1}{\gamma m_{e} }  
\label{53} 
\end{equation} 
for relativistic particles. However, from Section II, the average time between successive emissions of electrons and positrons directly evaporated by the black hole is
\begin{equation}
\Delta t_{e} \approx \frac{200}{E_{peak} } .
\label{54}
\end{equation}
Using $\gamma \approx E_{peak} /m_{e} $, we see that
\begin{equation} 
\Delta t_{c} << \Delta t_{e}  
\label{55} 
\end{equation} 
for all $T_{bh} >> m_{e} $. Thus the classical causality condition \eqref{53} is only satisfied by a negligible fraction of the emitted particles when $T_{bh} >> m_{e} $. The same conclusion applies for QCD particles when $T_{bh} >> \Lambda _{QCD} $. Because the emission of a given species is damped near its rest mass threshold \cite{P3}, so increasing $\Delta t_{e} $, the classical causality condition is unlikely to be satisfied even near particle rest mass thresholds. (Also recall that $\sigma _{brem} $ decreases faster with decreasing $E$ in the non-relativistic limit.) At any $T_{bh} $, if a photosphere is to form, we additionally require the causality constraint to be met within a sphere of radius $r_{brem} \approx 1/m_{e} $ around the black hole. This only further tightens the $\Delta t$ constraint.

Quantum mechanical effects relax the above classical constraint only slightly. By the Heisenberg Uncertainty Principle, the emission of each electron from the black hole is smeared out over a time $\Delta t\approx \left(\gamma m_{e} \right)^{-1} <<\Delta t_{e} $, leading to it having a spatial spread in the direction of emission of $\Delta r\approx \left(\gamma m_{e} \right)^{-1} $. Thus when the electron is a distance $r$ from the black hole, we can visualize it quantum mechanically as an eggshell segment of thickness $\left(\gamma m_{e} \right)^{-1} $, radial curvature $r$ and tangential width growing to  $1/m_{e} $, whose probability of overlapping another emitted electron is essentially given by the classical constraint above. Additionally, the virtual photon exchanged between the two bremsstrahlung-interacting electrons, which typically transfers 3-momentum $|\boldsymbol{k}_{\gamma *} |\approx m_{e} $ and negligible energy and has a nonzero mass ($m_{\gamma *} ^{2} \approx -m_{e} ^{2} $), cannot be localized within a length less than $1/|\boldsymbol{k}_{\gamma *} |\approx 1/m_{e} $. It can extend over a spacelike interval but spacelike intervals in the propagator are strongly suppressed. However, the wave speed at which the field of one electron propagates to the other electron in the Dirac equation remains $c$. Therefore the above classical causality constraint essentially always applies to the quantum interaction.

Equation ~\eqref{19} then must be replaced with an expression which more accurately describes the geometry of the interaction and accounts for the shortened histories of the interacting particles. As we discuss in Ref.~\cite{PCM}, the causality constraint and truncated histories also decrease the momentum transferred in each interaction, decreasing the effective $\sigma _{brem} $ well below the already small $\sigma _{brem} \propto \alpha ^{3} $ cross-section of Eq.~\eqref{17}. Even at high $T_{bh} $, the average number of interactions per particle is bounded by $\Delta t_{c} /\Delta t_{e} \approx 10^{-2} $ and cannot reach ${\mathcal N}=1$.  If an emitted particle does scatter once, we now show that the distance to complete that first scattering is such that subsequent complete scatterings by the particle within $r_{brem} $ of the black hole are not possible. Thus it is improbable for particles to experience more than one bremsstrahlung interaction. This by itself is sufficient to prevent QED photosphere formation.

\subsection{Scale for Bremsstrahlung Interaction}

 In the original Heckler model the distance needed for the completion of a bremsstrahlung interaction, i.e. the distance required for the formation of the final on-shell electron, is assumed to be $d_{form} \approx 1/m_{e} $ in the CM frame of the interaction. This is not correct and must be replaced by the much larger distance $d_{form} \approx E/m_{e} ^{2} $, as we can see in two ways.

 First, the 4-momentum of the off-shell electron is ${\rm p}_{{\rm e}*} ^{2} =m_{e} ^{2} +2{\rm p}_{{\rm 1}} ^{{'} } {\rm .k}$ where ${\rm p}_{{\rm 1}} ^{{'} } $ and ${\rm k}$ are the 4-momenta of the final on-shell electron and photon, respectively (see Fig. 1). From Eqs.~\eqref{32} and \eqref{33}, the average energies of the final electron and photon are $\sim E/2$ and the average angle between them is $\phi _{av} \approx m{}_{e} /2E$, so we have ${\rm p}_{{\rm e}*} ^{2} -m_{e} ^{2} \approx m_{e} ^{2} /2$. Thus the average lifetime of the off-shell electron in its own frame is $\sim 1/m_{e} $, which corresponds to $\sim E/m_{e} ^{2} $ in the CM frame of the interaction.

 Second, following Ref.~\cite{LPM}, consider the final electron and photon as they are being created. Because they are created with a small separation angle, their wave packets will initially overlap for some distance as they propagate. This distance is the formation distance. The transverse momentum components of the final electron and photon relative to the momentum of the initial electron $\boldsymbol{p}_{1} $ are both $p_{T} \approx \phi _{av} E$. By the Heisenberg Uncertainty Principle these transverse momenta can be resolved only when the transverse spatial separation is at least $1/p_{T} $. Since the opening angle between them is $\phi _{av} $, this occurs after they have traveled a radial distance $\left(p_{T} \phi _{av} \right)^{-1} \approx E/m_{e} ^{2} $ in the CM frame of the interaction. Thus again the formation distance must be $d_{form} \approx E/m_{e} ^{2} $.

 As we have already noted, the consequence of this is that the only relevant interactions for photosphere creation can be those which occur within $r_{brem} \approx 1/m_{e} $ of the black hole. Certainly $d_{form} >>1/m_{e} $ for $T_{bh} >>m_{e} $ and so, on completion of the first bremsstrahlung interaction, the products can no longer significantly interact with other particles. In the extremely rare event that an electron undergoes more than one scattering within $r_{brem} $ of the black hole, the subsequent scattering occurs when the electron is off-shell, not on-shell.

\subsection{Off-Shell Interactions and the LPM effect}

In the general setting, if an off-shell particle undergoes multiple scattering over the distance $d_{form} $ (the formation length of the photon and final electron in the bremsstrahlung interaction), Landau, Pomeranchuk and Migdal have shown that the bremsstrahlung cross-section is decreased by a numerical factor of up to $O\left(1\right)$ (see \cite{LPM0}; for a recent review see Ref.~\cite{KL}). This is known as the LPM effect and has been confirmed experimentally, at least in the soft photon regime \cite{SLAC}. (The effect is greatest in this regime because $d_{form} $ increases as  $\omega $ decreases \cite{LPM3}.) The suppression of the bremsstrahlung cross-section is due to multiple scattering breaking the coherence of the final state as the photon is forming \cite{LPM2}. The suppression increases with electron energy and the number of off-shell scatterings or, equivalently, the number density. The LPM effect modifies both QED and QCD bremsstrahlung. Recent theoretical treatments include elastic (Coulomb) and inelastic (bremsstrahlung) scatterings of the off-shell electron \cite{LPM2,LPM}.

 Applying the LPM effect to the black hole scenario, we would expect multiple scatterings of the off-shell electron to decrease the effective bremsstrahlung cross-section. Since the off-shell electron requires a distance $d_{form} $ to decay into the final on-shell electron and photon, even if ${\mathcal N}>1$ within $r_{brem} \approx 1/m_{e} $, each particle generally undergoes at most one complete on-shell scattering as it streams away from the hole. Recall that the probability of undergoing even one complete scattering is significantly damped by the causality constraint at $T_{bh} >>m_{e} $. Incorporating the precise effect in the black hole scenario requires calculating the LPM effect from first principles because the black hole scenario differs from other LPM scenarios in at least three ways: (i) the black hole off-shell interactions are bremsstrahlung (not just Coulomb scatterings) with average momentum transfer similar to that of the initial bremsstrahlung interaction; (ii) the off-shell interactions and the initial interaction occur essentially simultaneously, not sequentially, and their range is greater than the distance between each off-shell interaction; and (iii) the non-uniform number density, truncated particle histories and energy distributions of the emitted particles are specific to the black hole case.

 Proceeding qualitatively, we expect the following outcomes for black hole QED interactions. For bremsstrahlung, we have $r_{brem} \approx 1/m_{e} <<d_{form} $, so multiple scatterings, if they occur, can perturb only a small segment of the complete interaction of a typical particle. Because $d_{form} >>r_{brem} $, any scatterings affect the particle essentially simultaneously within a distance $r\lesssim r_{brem} $ of the black hole. (The actual region will be smaller than $r_{brem} $ because of the causality constraint.) Therefore the particle can be thought of as interacting with a region within $r\lesssim r_{brem} $ as a whole. This corresponds to the single radiator or thin target LPM limit \cite{FOM,WGP, KL, BK}. The probability for interaction in this limit is given by the total scattering angle within the $r\lesssim r_{brem} $ region \cite{KL}. If we approximate this angle by that appropriate for a random walk of successive scatterings, $\theta _{MS} \approx \sqrt{{\mathcal N}_{S} } \phi _{av} $ where ${\mathcal N}_{S} $ is the number of scatterings the particle undergoes within $r\lesssim r_{brem} $ of the black hole, then $\theta _{MS} \ge \phi _{av} $ for ${\mathcal N}_{S} \ge 1$. From the numerical simulations of Haug \cite{EH1}, $d\sigma _{brem} /d\theta $ falls off roughly exponentially at angles above $\phi _{av} $ and the momentum transfer is increased, although the energies of the final photon and electron stay of order $E$. Hence $d\sigma _{brem} \left(\theta _{MS} \right)/d\theta \le d\sigma _{brem} \left(\phi _{av} \right)/d\theta $ and the modified cross-section is always less than that for the interaction without off-shell scattering. Additionally, if the bremsstrahlung photon pair-creates before it is completely formed, $\sigma _{brem} $ is further suppressed \cite{KL}.

 Thus the net effect of multiple off-shell QED scatterings is to decrease the probability of any particle producing a bremsstrahlung photon. Even if ${\mathcal N}>1$, each particle requires a distance $d_{form} >>r_{brem} $ to complete the interaction and so is unlikely to undergo more than one completed bremsstrahlung interaction as it streams away from the black hole. Because each particle rarely undergoes more than one completed interaction, the number density around the black hole can increase by only a small fraction (at most 50\%) even if ${\mathcal N}>1$. Hence we conclude that a QED photosphere will not develop.

\subsection{The Plasma Mass Correction}

In the original Heckler model, it was argued that on scales less than $r_{brem} \approx 1/m_{e} $ the QED bremsstrahlung cross-section should be suppressed by Coulomb scattering off other electrons around the black hole. (However, Heckler did not include the off-shell bremsstrahlung interactions discussed above.) To account for the Coulomb scattering, Refs.~\cite{HE1,CL} took the simplified approach of defining the effective electron mass $m_{e} $ to include the fermion self-energy for a finite temperature bath. In the expression for $\sigma _{brem} $, $m_{e} $ was augmented in the interaction rest frame by the thermal plasma mass given by Eq.~\eqref{18}, which describes the screening of the electron by other emitted electrons. As Heckler pointed out, the Hawking emission is not a true plasma. Because the emitted particles flow isotropically out of the point-like black hole, the particle distribution cannot be transformed into a frame in which the particles exhibit a Maxwell-Boltzmann momentum distribution.

 As we have seen, the causality constraint prevents emitted electrons interacting significantly within $r_{brem} \approx 1/m_{e} $ of the black hole. This should apply to Coulomb scattering as well as bremsstrahlung interactions. Also, since each electron requires a distance $d_{form} >>r_{brem} $ to complete a bremsstrahlung interaction, $n\left(r\right)$ cannot increase substantially within $r_{brem} $ of the black hole due to interactions. Thus plasma mass considerations should be irrelevant within $r_{brem} \approx 1/m_{e} $ for $T_{bh} >>m_{e} $.

\subsection{Other Bremsstrahlung Processes}

 Although the Heckler photosphere scenario relies on the $O\left(\alpha ^{3} \right)$ two-body bremsstrahlung interaction between two charged particles emitted by the black hole, the emitted particles may experience other QED bremsstrahlung processes which we investigate in detail in the accompanying paper \cite{PCM}. First, there should be the $O\left(\alpha \right)$ inner bremsstrahlung generated by the charged particle as it escapes the black hole and, from the point of view of a distant observer, has its velocity changed from being essentially zero at the black hole to its asymptotic value. This inner bremsstrahlung emission will contribute about 10\% of the total photon power from an $M_{bh} =5\times 10^{14} {\rm \; g} $ black hole and has an approximately energy-independent power spectrum which dominates over the directly emitted Hawking photons at low energies \cite{PCM}. Secondly, there should be the $O\left(\alpha ^{2} \right)$ bremsstrahlung emission from the scattering of the charged emitted particles by the stochastic electromagnetic field of the black hole, and another $O\left(\alpha ^{3} \right)$ contribution from the scattering of the inner bremsstrahlung photon off an ambient charged particle. In fact, since the inner bremsstrahlung dominates all bremsstrahlung processes, the other bremsstrahlung processes can be simply regarded as small $O\left(\alpha \right)$ and $O\left(\alpha ^{2} \right)$ corrections to the inner bremsstrahlung.  The total bremsstrahlung emission is then essentially determined by the asymptotic momentum distribution of the charged particles, which to lowest order in $\alpha $ is determined by the Hawking emission formula Eq.~\eqref{1} and to higher orders is modified slightly by the interactions between the emitted charged particles. In all cases, however, the outgoing photon and electron from any bremsstrahlung interaction require a formation distance $d_{form} \approx \gamma /m_{e} $. Thus none of these bremsstrahlung interactions can produce a QED photosphere around a $T_{bh} >>m_{e} $ black hole.

\subsection{Extension to Pair-Production and QCD Effects}

 So far, we have discussed QED bremsstrahlung interactions but the above considerations also apply to QED pair production $e+\gamma \to e+e^{+} +e^{-} $. The pair-production cross-section is obtained from $\sigma _{brem} $ by rotating the Feynman diagrams to depict an incident photon rather than an outgoing photon and using the substitution rules \cite{EH1}. In the relativistic limit, $\sigma _{pair} $ has the same form as $\sigma _{brem} $ \cite{EH2}. The distribution of the outgoing $e^{+} $ and $e^{-} $ is strongly peaked in the direction of the initial photon and each carries on average approximately half the initial photon energy \cite{EH1}. The formation length of the $e^{+} e^{-} $ pair is also $d_{form} \approx \omega /2m_{e} ^{2} >>1/m_{e} $ when $T_{bh} >>1/m_{e} $ \cite{LPM3}. As before, any multiple off-shell scattering will suppress $\sigma _{pair} $ within the distance $d_{form} $ when $T_{bh} >>m_{e} $. Thus each particle can undergo at most one complete pair-production event as it streams away from the black hole and the probability of that interaction is strongly suppressed by the causality constraint. Again no QED photosphere can develop.

 The above discussion also applies to QCD gluon-bremsstrahlung and QCD pair-production interactions of the emitted particles. (Note here we are referring to the QCD interactions between particles in different QCD jets, not between particles in the same QCD jet. The interactions between particles in the same jet are included in the fragmentation and hadronization modeling of Ref.~\cite{MW}.) To derive the corresponding QCD quantities, $m_{e} $ is replaced by the quark mass $m_{q} $ or $\Lambda _{QCD} $, and $\alpha $ by the strong coupling constant $\alpha _{s} $. This leads to Eq.~\eqref{29} in the relativistic case, neglecting color factors of $O\left(1\right)$. The average time between QCD particle emissions by the black hole is $\Delta t_{QCD} \approx 20/E_{peak} $ when $T_{bh} >>\Lambda _{QCD} $ . The QCD formation scale $d_{form}^{QCD} \approx E/\Lambda _{QCD} ^{2} $ is also the scale at which the Hawking emission hadronizes. Applying the same arguments as above, when $T_{bh} >>\Lambda _{QCD} $, the causality constraint implies that the emitted QCD particles have little chance of interacting within $r_{brem}^{QCD} \approx 1/\Lambda _{QCD} $ of the black hole and any particles that do interact can undergo at most one complete QCD bremsstrahlung interaction within $d_{form}^{QCD} >>1/\Lambda _{QCD} $. Thus no QCD photosphere (also known as a chromosphere) should form when $T_{bh} >>\Lambda _{QCD} $. Because of the time separation between subsequent QCD jets and the LPM effect, nor should a photosphere form from interaction of the hadronization products which appear at $d_{form}^{QCD} $.

QCD phenomenology and experiments also strongly argue against a QCD photosphere developing when $T_{bh} $ (or more specifically the peak energy of the directly emitted flux) passes through values that are roughly $\Lambda _{QCD} $. Because the Heckler estimate, $T_{crit}^{QCD} \approx \Lambda _{QCD} /\alpha _{s} ^{5/2} $, is close enough to $\Lambda _{QCD} $ that high energy formulae may be inappropriate, the question arises of whether a QCD photosphere could form from \textit{non-relativistic} effects at these temperatures. If the emission were non-relativistic, the causality constraint would be weakened. Also at low energies the effective $\alpha _{s} $ is observed to increase from its high energy limit,  $\alpha _{s} =12\pi /\left\{\left(33-2n_{f} \right)\ln \left(E^{2} /\Lambda _{QCD} ^{2} \right)\right\}$ where $n_{f} $ is the number of relevant quark species, and may approach $\pi $ as $E\to \Lambda _{QCD} $ \cite{SCH,AMN,BCLM,DEU}. On the other hand, it is not clear whether, even if multiple scatterings were to occur around $T_{bh} \sim \Lambda _{QCD} $, they would affect the observable spectra, because the Hawking emission is strongly damped and the multiplicity per QCD jet is very low around the $\Lambda _{QCD} $ threshold \cite{MW}. However, we deduce from both the QCD phenomenology and experimental perspectives that when the black hole begins to emit asymptotically free quarks, these quarks are relativistic. Therefore no QCD photosphere should develop around $T_{bh} \sim \Lambda _{QCD} $.

The emission picture which is consistent with QCD phenomenology and threshold accelerator experiments is as follows. When $T_{bh} $ (or more precisely $E_{peak} \approx 2.81T_{bh} $ for $s=0$) is about the pion rest mass ($m_{\pi ^{\pm } } \approx 140{\rm \; MeV}$ and $m_{\pi ^{0} } \approx 135{\rm \; MeV}$), the black hole should start emitting pions directly, with the grey-body factors and degrees of freedom appropriate for massive $s=0$ particles modified by the finite-size structure of the pion. Once the scale of the black hole emitting region becomes small compared with the length below which the constituent (valence) $u$ and $d$ quarks in the pions become asymptotically free, the black hole should begin emitting the quarks individually. These quarks then hadronize into pions on the proper distance scale $1/\Lambda _{QCD} $ in the frame of the emitted quark. The energy at which this happens is sufficiently above the pion rest mass that the pion is relativistic. Hence the pion's constituent quarks should also be regarded as relativistic. Furthermore, during the asymptotic freedom regime before the quark hadronizes, the quark masses which are used in the QCD Lagrangian are the much lighter quark current masses, not the quark constituent masses. (The quark current masses are $m_{u} \approx (1.5-3){\rm \; MeV}$, $m_{d} \approx (3-7){\rm \; MeV}$ and $m_{s} \approx (70-120){\rm \; MeV}$, whereas the quark constituent masses are $m_{u,d} \approx 0.31{\rm \; GeV}$ and $m_{s} \approx 0.48{\rm \; GeV}$. For the heavy $c$, $b$ and $t$ quarks, the current masses are approximately equal to the constituent masses and lie above ${\rm 1\; GeV}$ \cite{PDG}. The quark constituent masses are the relevant masses as the quarks hadronize and afterward.) Thus the quarks are ultra-relativistic during their asymptotic freedom phase and then gain larger effective masses, but still remain relativistic, when they hadronize.

The hadronization process itself at these energies also cannot produce a significant increase in the number of final particle states. What the quarks hadronize into is dictated by energy and quantum conservation laws. Hence, around $T_{bh} \sim \Lambda _{QCD} $, the quarks can hadronize only into pions (or the few light meson states which decay into pions or mimic $\pi ^{0} $ decay into photons) and the multiplicity of the final pions per initial Hawking emitted quark must be very low (one or two pions per initial quark) \cite{PDG}. ~Emission of $e^{+} e^{-} $ or photons by these quarks or pions is covered by our earlier analysis of QED interactions and so is negligible, as are any weak interactions. Gluon bremsstrahlung by quarks at these energies is also inconsequential. The gluon has to obey the QCD confinement (no-free-color) and quantum conservation rules. Thus a gluon must have an energy of about $\Lambda _{QCD} $ or greater to produce free final state particles, the lowest energy option being a pion pair. Because the total momentum in the frame defined by the black hole and the Hawking emitted quark must be conserved and there are no available very light final states for the gluon or the quark, an asymptotically free Hawking emitted quark of energy $E\sim \Lambda _{QCD} $ cannot be slowed down by gluon bremsstrahlung, except by the gluons involved when that quark eventually hadronizes.

The above description is completely consistent with accelerator experiments. The cross-section for $e^{+} +e^{-} \to \; pions$ has been measured at center-of-mass energies from close to the $2m_{\pi } $ threshold to 1 GeV and remains smooth over that region (see Ref.~\cite{DEHZ} for a recent review). No `dip' or other structure in the data is observed around center-of-mass energies of $\Lambda _{QCD} $ where the behavior changes from direct production of pions to the production of final state pions via an intermediate asymptotically free quark state.

To further visualize the black hole quark emission and hadronization process around $T_{bh} \sim \Lambda _{QCD} $ and at higher temperatures, one can use the chromoelectric flux-tube model (also known as the QCD string model) that successfully describes hadronization in accelerators (see \cite{GG}, \cite{Nam}; for recent reviews see Refs.~\cite{BV,Stz}). In this interpretation a quark is emitted from the black hole dragging behind it a tube of color field whose other end is attached to the black hole. The string tension $\sigma _{\Lambda } $ of the flux-tube is constant, so that its potential energy $\sigma _{\Lambda } x$ grows linearly with the length of the tube $x$. When the flux-tube has been stretched to the length $x\approx \sqrt{2\pi /\sigma _{\Lambda } } \approx 1/\Lambda _{QCD} $ at which the potential energy matches the energy required to create a new $q\bar{q}$ pair out of the vacuum (including the transverse momentum of the $q\bar{q}$ pair implied by the Heisenberg Uncertainty Principle applied to the flux-tube width), the tube breaks in two, producing a $q$ and a $\bar{q}$ attached to the two new ends where the tube broke \cite{Stz}. To conserve energy, the stretching and  $q\bar{q}$ creation is accompanied by a corresponding deceleration of the lead quark. This stretching and breaking sequence continues until the energy of the lead quark has been expended producing new $q\bar{q}$ pairs. The quarks and antiquarks created from the flux-tube match up to hadronize into bound color-neutral states. The whole process is completed over the distance scale $E/\Lambda _{QCD} ^{2} $ in the black hole frame, where $E$  is the energy of the initial quark. The flux-tube model naturally gives a multiplicity growth of final states per initial quark that is proportional to $\ln E$, as observed in accelerators.

That the asymptotically free quarks are relativistic, rather than non-relativistic, when first emitted by the black hole around $T_{bh} \sim \Lambda _{QCD} $ can additionally be seen from the thermodynamical `detailed balance' derivation of the Hawking flux. When a black hole is in quasithermal equilibrium at a given instant with a surrounding thermal bath of the same temperature, the absorption rate of a given species equals its emission rate. If a pion incident upon a black hole has total energy insufficient for the pion to break up into subparticles with one subparticle going down the hole and one escaping, detailed balance gives the Hawking emission rate of direct outgoing pions from the capture cross-section of the incoming pions.  When the pion size is comparable to the cross-section for a point-like scalar particle of the same mass and energy as the pion, the pion capture cross-section should be similar to that of the corresponding point-like scalar particle.  (This cross-section should also apply when the black hole itself is much smaller than the pion if the pion is moving sufficiently slowly that the capture cross-section is much larger than the hole itself.  In this case, once the pion falls to a distance from the hole comparable to its size, it will be nearly inevitable that part of the pion will fall into the hole, with the rest having insufficient energy to escape even though the pion may initially appear to have too large an intrinsic size to be captured by the smaller hole.) If the black hole size is significantly smaller than the size of the pion and the incoming pion is moving relativistically (that is, its kinetic energy is significantly greater than its rest mass), the black hole can capture an individual constituent quark of the pion, with the other constituent escaping capture or being captured after a significant time interval. This inelastic scattering of the pion by the black hole could also lead to the production of other particles, depending on the available energy. One could visualize the capture process as the glue string between the captured quark and the constituent antiquark of the original pion breaking in two, with a new antiquark forming on the end of the string attached to the quark that fell into the hole and a new quark forming at the end of the string attached to the uncaptured antiquark.  The new antiquark attached to the glue string going into the hole could be captured by the hole, while the new quark attached to the external glue string could escape along with the original antiquark.  For incident pions of high enough energy, more complicated inelastic processes could lead to greater multiplicity and number of species in the escaping and captured states. The reverse process of the absorption of constituent quarks corresponds to the Hawking emission of individual relativistic quarks that hadronize into one or more pions and/or other mesons and baryons.

From our above discussion it is clear that the QCD production around a $T_{bh} \sim \Lambda _{QCD} $ black hole is not analogous to the creation of quark-gluon plasma at the Relativistic Heavy Ion Collider (RHIC) (for a recent review of RHIC results, see Ref.~\cite{MN}). At RHIC the collision of heavy ${}^{197} {\rm Au}+{}^{197} {\rm Au}$ nuclei at center-of-mass energies of $130 {\rm \; GeV}$ and $200 {\rm \; GeV}$ per nucleon generates a strongly coupled, high entropy quark-gluon plasma which rapidly achieves a thermal spectrum and then expands isentropically until the particles hadronize and decouple by a time $10 - 20 {\rm \; fm/c}$. Presently it is not known whether the plasma achieves its thermal spectrum of states by interactions or is initially created with that distribution of states \cite{Stz}. At the moment of collision, the overlapping ${}^{197} {\rm Au}$ nuclei are highly Lorentz-contracted in the center-of-momentum frame, implying an overlap time of $0.2 {\rm \; fm/c}$ and an initial thermalized energy density of about $15-3000{\rm \; GeV\; fm}^{-3} $. The initial energy density estimate varies with the interpretative model but in all cases is far greater than the $1{\rm \; GeV\; fm}^{-3} $ threshold for quark-gluon plasma formation or the density within a nucleon ${\rm 0.5\; GeV\; fm}^{-3} $, and remains so for up to $5 {\rm \; fm/c}$. The number of charged particles arriving at the detectors (about $5000$) indicates that approximately $20$ high energy particles are created per initial nucleon but the dependence of the charged particle multiplicity on the available energy per colliding nucleon still matches that seen in $e^{+} +e^{-} $ and $p+p$ accelerator collisions \cite{BBB}. The hadronic multiplicities are determined by the nonzero baryonic chemical potential (baryon/anti-baryon asymmetry) of the initial nuclei. Notably, a reduction of a factor of about $5$ in the number of high transverse momentum hadrons is observed. This phenomenon is known as jet-quenching, although the high transverse momentum jets are not truly quenched but spread over a wider angle with lower average energy per particle. The hadronic energy loss implies significant QCD interaction with the ambient plasma, but photons are observed to travel through the plasma unimpeded. The jet-quenching can be modeled by assuming that the collision initially creates a gluon number distribution of $dN_{g} /dy\sim 1000$, where $y$ is the rapidity defined by $E\equiv \sqrt{m^{2} +p_{T} ^{2} } \cosh y$ for a particle of energy $E=\gamma m$ and transverse momentum $p_{T} $. (Integrating over $1\le \cosh y\le \gamma $, this gives the number of gluons per nucleon to be of order $10 - 10^{2}$). Application of the Heisenberg Uncertainty Principle indicates that the nuclei are gluon-saturated on the scale of the collision even before impact: that is, at 200 GeV per nucleon the two gold nuclei see each other as dense objects, not ensembles of distinct partons, because the atomic number (parton density) and transverse momenta are such that the partons overlap considerably in the transverse plane \cite{DKN}. All of the above effects attributable to a strongly-coupled quark-gluon plasma are experimentally confirmed to be suppressed at lower collision energies and in deuteron-gold collisions \cite{DE}.

 In the $T_{bh} \sim \Lambda _{QCD} $ black hole case, however, in contrast to RHIC, the conditions for the simultaneous production of an ultra-high density of QCD particles are not met. A $T_{bh} \sim \Lambda _{QCD} $ black hole emits particles by the Hawking process sequentially and relativistically with significant time separation between each particle emission. Thus, as we have seen in Section IV.F, negligibly few of the Hawking emitted quarks are in sufficient causal contact to interact with each other within $1/\Lambda _{QCD} $ of the black hole. Furthermore, as discussed above, conservation of energy severely limits the number of final states that can be created per initial quark around $T_{bh} \sim \Lambda _{QCD} $. In the black hole case, only an energy $E\sim \Lambda _{QCD} $ is available per Hawking emitted quark, in contrast to the $200 {\rm \; GeV}$ per nucleon available at RHIC. (Additional differences include the facts that the corresponding baryonic chemical potential is $\mu _{B} =0$, because the black hole emits equal numbers of quarks and antiquarks, and that the black hole particles are not initially gluon-saturated.) Hence, as given by Eqs.~\eqref{11} -- \eqref{15}, the greatest achievable density around a $T_{bh} \sim \Lambda _{QCD} $ black hole is well below $1{\rm \; GeV\; fm}^{-3} $. The closest experimental analogy to the $T_{bh} \sim \Lambda _{QCD} $ black hole situation is the generation of pions and QCD jets in $e^{+} +e^{-} $ accelerator collisions and in hadronic collisions at much lower center-of-mass energies per nucleon.

In summary we conclude that the time separation between emissions and the limited average energy available per Hawking emitted particle prevent the formation of a QCD photosphere around a $T_{bh} \sim \Lambda _{QCD} $ black hole, and the causality and $d_{form}^{QCD} >>1/\Lambda _{QCD} $ constraints prevent QCD photosphere formation at $T_{bh} >>\Lambda _{QCD} $. We also note that the above description of the QCD threshold emission processes should lead to more precise numerical modeling of the astrophysical particle spectra produced by Hawking radiation around the $\Lambda _{QCD} $ threshold than has previously been published.

\subsection{Comments on Cline, Mostoslavsky, and Servant Simulation of Heckler Photosphere}

 The original Heckler scenario was numerically modelled by J. Cline, Mostoslavsky, and Servant \cite{CL}, who simulated photosphere development using the test particle method to solve the Boltzmann equation for propagation of plasma particles undergoing collisions \cite{CL}. They found that QED photosphere production set in at $T_{crit} \approx 50{\rm \; GeV}$ (comparable to Heckler's value of $45 {\rm \; GeV}$) but extended over a much smaller region: it started at $r_{in} \approx 10^{4} r_{bh} $ (compared to Heckler's $r_{in} \approx 10^{9} r_{bh} $) and ended at $r_{out} \approx 1/m_{e} $ where the trajectories of particles within the interaction range rapidly become parallel, as we saw in Section IV.D. For $T_{bh}\gtrsim 50 {\rm \; TeV}$, their photosphere region was extended by Compton scattering effects. However, Cline, Mostoslavsky, and Servant use $\sigma _{C} =2\pi \alpha ^{2} \ln (E/m_{e} )/(m_{e} E)$, whereas the correct relativistic Klein-Nishina cross-section \cite{KN} for a photon of energy $E$ Compton scattering off an electron of energy $E$ in the black hole rest frame is $\sigma _{C} \approx 2\pi \alpha ^{2} \ln (E/m_{e} )/E^2$. They also found that the fluid description is of questionable validity and that it led to an average photon energy at the outer photosphere edge which differed from Heckler's value, even decreasing as $T_{bh} $ increases. The differences are more marked in the QCD case, partly because of the form of the QCD plasma mass correction used by Heckler. They found that a QCD photosphere forms for $T_{bh} \gtrsim T_{crit}^{QCD} \approx 175 {\rm \; MeV}$ and extends to $r_{out} \approx 1/\Lambda _{QCD} $.

The Cline, Mostoslavsky, and Servant work is elucidating but, because it takes the same approach as Ref.~\cite{HE1}, all of the points we have made above apply to it. Most importantly, the authors neglect the causality and $d_{form} $ constraints which limit each particle to at most one on-shell bremsstrahlung interaction within $r_{out} \approx 1/m_{e} $ or $r_{out} \approx 1/\Lambda _{QCD} $ and make even that highly unlikely. An additional problem is that they generate the bremsstrahlung photons with random energy $0\le \omega \le E$ and random angle $0\le \phi \le 2\pi $. We have seen that the true photon angle is very small $\phi _{av} \approx m_{e} /2E$ and deviates little from the direction of the initial electron. Randomizing the angle would exaggerate photosphere development, and this may explain why Cline, Mostoslavsky, and Servant find  $T_{crit} $ to be lower than Heckler's original analytic estimate. Their numerical method should therefore be modified to include all of our previous points in Section IV and could be employed to investigate the consequences of the rare bremsstrahlung interactions.

\subsection{Comments on Kapusta and Daghigh Models}

 Kapusta \cite{Kap1,Kap} and Daghigh and Kapusta \cite{DK1,DK2,DK3} have analytically and numerically investigated the hydrodynamic expansion of matter around a black hole in the original Heckler photosphere formation scenario. In their treatment, the $T_{bh} >100{\rm \; GeV}$ emission is assumed to be a quasi-stationary shell of plasma expanding relativistically away from the hole. The plasma is kept in local thermal equilibrium by the QED and QCD bremsstrahlung and pair-production interactions of the original Heckler model. `Local thermal equilibrium' in this context is taken to mean that the mean-free-path, or thermalizing scale, of the plasma particles is less than the scale over which the plasma temperature is changing \cite{DK1}. They find that the quasi-stationary photosphere shell extends out to $10^{-10} \left(T_{bh} /1{\rm \; TeV}\right)^{-2} {\rm \; cm}$. The expected observable photon signal is then calculated from the luminosity of the photosphere surface.

 The results of Daghigh and Kapusta are not supported by our analysis. As we have seen in Section IV, the interactions within $r\lesssim r_{brem} $ of the black hole cannot significantly increase the number density $n\left(r\right)$, as required for photosphere formation. Their approach also neglects the causality and $d_{form} $ constraints within $r\lesssim r_{brem} $.  Once a particle reaches $r_{brem} $, we have shown that the geometry is such that it has little probability of interacting with other emitted particles. Thus, contrary to the claim of Daghigh and Kapusta, there should be no change in the particle energy after the particle reaches $r\approx r_{brem} $ and no persistent QED or QCD photosphere or locally thermalized plasma should develop.

\section{CONSEQUENCES FOR OBSERVABLE SPECTRA}

In Section IV we have shown that a Heckler QED photosphere will not form around a Hawking black hole. Similar arguments and additional QCD phenomenological considerations prevent a QCD photosphere developing. Interactions which occur within $r_{brem} $ of the black hole determine whether a Heckler photosphere can arise. These interactions are extremely rare and any multiple scatterings involve off-shell particles. In contrast, the original Heckler model postulated that each particle on average undergoes successive on-shell bremsstrahlung interactions out to a much greater radius. This led to an increase in the expected flux at low energies and a steepening of the spectra at high energies. Including the off-shell scatterings, however, suppresses the effective bremsstrahlung cross-section and permits at most one completed bremsstrahlung interaction per particle. This one net interaction can only slightly modify the observable emission spectra, and not significantly as would a photosphere. Our remarks apply equally to pair production. Also of crucial importance to determining the full spectra is understanding the emission processes around $T_{bh} \sim \Lambda _{QCD} $.

 If each emitted particle produces at most one bremsstrahlung particle or undergoes pair-production once as it streams away from the black hole, there will be little effect on the detectability of an individual black hole or a background of black holes. The number of relativistically emitted particles of the relevant species in the relevant part of the spectra could be increased by at most a factor of $3/2$ and the average energy decreased by at most $1/3$. Furthermore, given the low likelihood of even one net interaction, the causality constraint discussed in Section IV.F implies that any change to the spectra should be several orders of magnitude weaker than this. As we show in Ref.~\cite{PCM}, inclusion of the causality constraint in an impact-parameter-based analysis of the two-body bremsstrahlung interaction around a $T_{bh} >>m_{e} $ black hole leads to an upper limit of $10^{-7} $ on the total fraction of energy emitted in two-body bremsstrahlung photons.  Charged and/or colored particles ($e^{\pm } $, $\mu ^{\pm } $, $\tau ^{\pm } $, $q^{\pm } $, $g$) and photons which are directly emitted by the black hole could participate in the rare bremsstrahlung or pair-production interactions. Additionally, almost all of the $\gamma $, $p$, $\overline{p}$, $e^{+} $, $e^{-} $, $\nu $ and $\overline{\nu }$ flux from a $T_{bh} \gtrsim 1 {\rm \; GeV}$ black hole comes from the decays of directly emitted particles. Thus all observable spectra, except the highest energy $\nu \overline{\nu }$ flux which is directly emitted, may be very slightly modified by interactions. As we discuss in Ref.~\cite{PCM}, at low energies the photon spectrum from an individual black hole will be enhanced by two other $O\left(\alpha \right)$ and $O\left(\alpha ^{2} \right)$ bremsstrahlung processes. In contrast, the low energy particles in the spectra from a PBH background were originally emitted with higher energies and then cosmologically redshifted to low energies \cite{JHM}, and so the background spectra can be only slightly modified at low energies (by the rare interactions near the black hole). In summary, the observable spectra from an individual black hole or PBH background remain those of the standard Hawking picture. The time-integrated background spectra above $E\approx 300$ MeV should maintain their original $dN/dE\propto E^{-3} $ form \cite{MC} with little amplitude change, in contrast to the $E^{-4} $ slope predicted by the Heckler model \cite{HE2}. The limits on $\Omega _{pbh} $ derived at $E\approx 100{\rm \; MeV}$ using the $\gamma $, $p$, $\overline{p}$, $e^{+} $, $e^{-} $, $\nu $ and $\overline{\nu }$ background spectra \cite{MC} will not be significantly affected.

We also note that the high energy limits on the detectability of individual holes, which are derived at $E>>\Lambda _{QCD} $, are not sensitive to behaviour around $T_{bh} \sim \Lambda _{QCD} $ but the limits on $\Omega _{pbh} $, which are derived at $E\approx 100 {\rm \; MeV}$, may be. This is because at $E\approx 100 {\rm \; MeV}$  most particles in the integrated spectra from a PBH background are QCD jet products from $T_{bh} \gtrsim \Lambda _{QCD} $ black holes emitting in the present era \cite{JHM}. The Galactic antiproton limit is less influenced by the behaviour around $T_{bh} \sim \Lambda _{QCD} $ because most of the postgalactic PBH antiprotons are generated in somewhat higher energy QCD jets, due to the greater rest mass of the antiproton \cite{JHM}. In future work a more precise numerical modeling of the spectra at $E\approx 100$ MeV using our insights of Section IV.K should be pursued.

\section{OTHER PHOTOSPHERE SCENARIOS}

\subsection{Belyanin, Kocharovsky, and Kocharovsky Model}

 Belyanin, Kocharovsky, and Kocharovsky \cite{BY} have proposed an unrelated photosphere model, in which charged particles from the black hole --- predominantly pions from the decay of Hawking radiated quarks and gluons --- form a well-defined `collisionless' plasma at $r>>r_{bh} $. The hydrodynamical behavior of this plasma is assumed to be maintained by collective wave-particle interactions and/or a self-induced turbulent magnetic field, both of which the authors postulate can arise from anisotropies in the charged particle distributions. Particle-particle collisions themselves are insufficient to maintain this hydrodynamical regime. Belyanin, Kocharovsky, and Kocharovsky claim that there is sufficient time for plasma wave turbulence to develop as the particles stream away from a high temperature black hole. Explicitly, they find that, when $T_{bh} \gtrsim 10{\rm \; GeV}$, the size of the plasma (taken to be the distance from the black hole) is greater than the Debye radius $r_{D} $ (effectively the inverse of the Langmuir oscillation frequency --- or plasma frequency --- generated by the charged particle separation), and that $r_{D} $ is greater than the average separation between charged particles.

The authors apply magneto-hydrodynamical equations for a plasma magnetic field in the case where the magnetic field is maximal, i.e. where the energy is equipartitioned between particle kinetic energy and the magnetic field. They argue that, for $T_{bh} \gtrsim 10 {\rm \; GeV}$, synchrotron radiation and electromagnetic cascading in the self-induced magnetic field will produce an observable MeV $\gamma$-ray burst lasting $10^{-1} - 10^{3} {\rm \; s} $. They envision three possible sources for the induced turbulent magnetic field. These involve the development, as the charged particles move away from the black hole, of unspecified plasma-wave turbulence or strong internal shocks. Suggested shock generators are $e^{+} e^{-} $ pairs created by electromagnetic cascade in the postulated magnetic field of the emitted particles, or collisions of emitted particles with the interstellar medium.

The Belyanin, Kocharovsky, and Kocharovsky scenario makes the following assumptions: that the charged particles interact sufficiently to form a plasma; that a substantial magnetic field is generated by anisotropies in the charged particle distributions; and that significant turbulence arises in this field. In analyzing this scenario, we find that a number of the model components have been treated incorrectly. We conclude that a self-induced photosphere of this type should not occur around the black hole.

 First, Langmuir oscillations are longitudinal oscillations that arise in a plasma when the positively charged particles move collectively in the opposite direction to or independently of the negatively charged particles. The mutual attraction between the separated positive and negative charges provides a restoring force that constrains the charged particles to oscillate around their equilibrium positions in the rest frame of the plasma. The oscillations can propagate along this axis of particle motion. For example, Langmuir oscillations can occur in an ionic plasma placed in an external magnetic field. However, in the case of particles streaming away from a black hole, we note that there is no initial asymmetry in the distribution or response of positive and negative charges --- for every particle, a neutral black hole stochastically emits on a short timescale an antiparticle of equal mass and opposite electric charge \cite{P1,P3}. Any initial black hole charge is lost on a timescale much less than the age of the Universe for $M_{bh} < 10^{5} M_{\odot } $ \cite{Z,G,CA,P1,P3}. The remaining random charge fluctuations are of order the Planck charge  $\left(\hbar c\right)^{1/2} =11.7e$ and therefore negligible \cite{P1,P3}. Additionally, a neutral black hole has no intrinsic magnetic field. If emitted particles decay, total charge symmetry is preserved. In particular, the net electric charge of the final products of a QCD jet initiated by a decaying quark is equal to the negligibly small charge of the original quark, i.e. $\pm e/3$ or $\pm 2e/3$. Moreover, self-induced QED turbulence is not observed in accelerator jets. Hence the outflow from a black hole remains essentially electrically neutral on all scales and Langmuir oscillations should not arise from charge asymmetries in the flux distribution.

 Although we do not expect a self-induced magnetic field to develop as the emitted particles stream away from the black hole, let us for the moment assume that collective wave-particle interactions and/or a turbulent magnetic field do arise in the charged particle flux when the hole is placed in an external magnetic field of sufficient strength and/or turbulence. Pions form via hadronization in QCD jets at a distance from the black hole of roughly
\begin{equation}
r_{\pi i} \approx \frac{\hbar c}{\Lambda _{QCD} } \left(\frac{5T_{bh} }{m_{q} } \right)\approx 10^{-12} \left(\frac{T_{bh} }{{\rm GeV}} \right){\rm \; cm},
\label{56}
\end{equation}
taking $m_{q} \approx 0.3 {\rm \; GeV}$ for $u$ and $d$ quarks, with faster pions forming farther from the black hole and more coaxially than slower ones. The charged pions will decay at a distance
\begin{equation} 
r_{\pi f} \approx \frac{E_{\pi } \tau _{\pi \pm ^{} } }{m_{\pi } c} \approx 10^{3} \gamma _{\pi } {\rm \; cm} 
\label{57}
\end{equation} 
where $E_{\pi } =\gamma _{\pi } m_{\pi } c^{2} $ is the energy of the charged pion and $\tau _{\pi \pm ^{} } =2.6 \times 10^{-8} {\rm \; s}$ is the charged pion lifetime. The average energy of the pions produced by the black hole is \cite{MW}
\begin{equation} 
\bar{E}_{\pi } \approx 0.6\sqrt{T_{bh} /{\rm GeV}} {\rm \; GeV}. 
\label{58} 
\end{equation} 
This value is distorted by the high energy $E_{\pi } \approx T$ particles but most pions in the instantaneous number flux are `slow' with energies $E_{\pi } \approx \left(1-3\right)m_{\pi } c^{2} $ \cite{MW}.

If Langmuir oscillations do arise in the charged pion flux, their frequency in the rest frame of the pion plasma is 
\begin{equation} 
\omega '_{\pi } =\left(\frac{4\pi e^{2} n'_{\pi } }{m_{\pi } } \right)^{1/2}  
\label{59} 
\end{equation} 
where the prime indicates quantities in the plasma rest frame, $e=4.8 \times 10^{-10} {\rm \; esu}$ is the electron charge in electrostatic units and $n'_{\pi } $ is the charged pion number density. Transforming into the rest frame of the black hole, the Langmuir frequency is
\begin{equation} 
\omega _{\pi } =\left(\frac{4\pi e^{2} n_{\pi } }{\gamma _{\pi } ^{3} m_{\pi } } \right)^{1/2}  \label{60} 
\end{equation} 
(see, for example, Ref.~\cite{BENB}). This disagrees with Eq.~(10) of Belyanin, Kocharovsky, and Kocharovsky which omits a factor of $\gamma _{\pi } ^{-1} $. In the black hole frame, the pion number density and flux are $n_{\pi } \approx \dot{N}_{\pi \pm ^{} } /4\pi r^{2} c$ and  $\dot{N}_{\pi ^{\pm } } \approx 10^{24} \left(T_{bh} /{\rm GeV}\right)^{3/2} {\rm \; s}^{-1} $, respectively \cite{MW}.

In the case of the slow pions with $\gamma _{\pi } \approx 1-3$, the first constraint for the hydrodynamical regime --- that the Debye radius $r_{D\pi } \sim c/\omega _{\pi } $ (which corresponds to $r'_{D\pi } \sim c/\omega '_{\pi } $, the distance at which the kinetic energy of a plasma particle balances its electrostatic potential energy in the plasma rest frame) be less than the characteristic size of the plasma $r$ --- is satisfied when $T_{bh}\gtrsim 1$ GeV. For the fast pions with $E_{\pi } \approx \bar{E}_{\pi } $, however, $r_{D\pi } $ is an order of magnitude greater than $r$ at all $T_{bh} $ and this constraint is never satisfied. In fact, at any given $T_{bh} $, the $r_{D\pi } \lesssim r$ constraint is satisfied only by pions with energy $E_{\pi } \lesssim 0.1\left(T_{bh} /{\rm GeV}\right)^{1/2} $ GeV. Thus the claim by Belyanin, Kocharovsky, and Kocharovsky that $r_{D\pi } \lesssim r$ is satisfied by all pions when $T_{bh} $ is greater than a few GeV is not supported by our analysis.

 The second constraint for the hydrodynamical regime --- that the Debye radius be greater than the average separation between charged pions in the plasma rest frame, i.e. ${r'}_{D\pi } > {n'}_{\pi } ^{-1/3} $ --- corresponds to
\begin{equation} 
r_{D\pi } >\gamma _{\pi } ^{4/3} n_{\pi } ^{-1/3}  
\label{61} 
\end{equation} 
in the black hole rest frame. The $\gamma _{\pi } ^{4/3} $ factor is omitted by Belyanin, Kocharovsky, and Kocharovsky. In the case of the slow pions, this constraint is satisfied when the pions reach a distance from the black hole of
\begin{equation}
r_{n\pi } \approx 10^{-14} \gamma _{\pi } ^{4} \left(\frac{T_{bh} }{{\rm GeV}} \right)^{{\rm 3/4}} {\rm \; cm}.
\label{62}
\end{equation}
In the case of the fast pions, it is satisfied when the pions reach a distance
\begin{equation}
r_{n\pi } \approx 10^{-15} \left(\frac{T_{bh} }{{\rm GeV}} \right)^{{\rm 1/2}} {\rm \; cm}.   \label{63}
\end{equation}
Recall, however, that the fast pions never satisfy the $r_{D\pi } <r$ constraint at any $r$. For all pions, we also have the third requirement --- that the first and second constraints can be satisfied only if $r$ is greater than the distance at which the pions form. Thus the region in which all three constraints are satisfied is ${\rm max}(r_{\pi i} ,r_{n\pi } ) \lesssim r \lesssim r_{\pi f} $ for slow pions and non-existent for fast pions.

 Repeating the above analysis for electrons and positrons, representative energies are $\bar{E}_{e} \approx 0.3\sqrt{T_{bh} /{\rm GeV}} {\rm \; GeV}$ for the fast $e^{\pm } $ and $E_{e} \approx 130m_{e} c^{2} $ for the slow $e^{\pm } $ \cite{MW}. The total $e^{\pm } $ flux produced by a black hole is  $\dot{N}_{e^{\pm } } \approx 10^{24} \left(T_{bh} /{\rm GeV}\right)^{3/2} {\rm \; s}^{-1} $ \cite{MW}.  For the slow $e^{\pm } $, it follows that $r_{De} \lesssim r$ when $T_{bh} \gtrsim 10^{3} {\rm \; GeV}$. For the fast $e^{\pm } $,  $r_{De} <r$ is never satisfied at any $T_{bh} $ and the $r_{De} \lesssim r$  constraint is satisfied only by those $e^{\pm } $ with energy $E_{e} \lesssim 10^{-3} \left(T_{bh} /{\rm GeV}\right)^{1/2} {\rm \; GeV}$. The second constraint, $r_{De} >\gamma _{e} ^{4/3} n_{e} ^{-1/3} $, is satisfied when the slow $e^{\pm } $ have reached a distance  $r_{ne} \gtrsim 10^{-12} \left(T_{bh} /{\rm GeV}\right)^{1/4} {\rm \; cm}$ from the black hole and when the fast $e^{\pm } $ have reached a distance  $r_{ne} \gtrsim 10^{-12} \left(T_{bh} /{\rm GeV}\right)^{3/2} {\rm \; cm}$. However, the $e^{\pm } $ produced by the $\pi ^{\pm } \to \mu ^{\pm } \to e^{\pm } $ decay chain form at a distance $r_{ei} \approx 10^{5} \gamma _{\pi } {\rm \; cm} $, which is much greater than $r_{ne} $. Combining these results, the region in which all three constraints are satisfied is $r_{ei} \lesssim r$ for slow $e^{\pm } $. For fast $e^{\pm } $, there is no region in which all three constraints are satisfied.

 In the above discussion, we assumed that the strength of the ambient magnetic field was sufficient to generate hydrodynamical behavior. Let us now estimate the strength of the external magnetic field required to cause significant longitudinal separation of positive and negatives charges as they move radially away from the black hole. As we saw in Eq.~\eqref{37}, the predominantly radially-moving particles may have a transverse velocity component $v_{T} \approx r_{bh} c/r$ in the black hole frame. Using the force equations $\boldsymbol{F}=\pm e \, \boldsymbol{v}_{T} \boldsymbol{ \times B}$ and $\boldsymbol{F}=\gamma ^{3} m \, {\rm d}\boldsymbol{v}/{\rm d}t$, the separation along the radial direction induced by the magnetic field is given by $\Delta r_{B} /r\approx eBr_{bh} /\gamma ^{3} mc$, implying
\begin{equation}
\frac{\Delta r_{B} }{r} \approx \frac{10^{-26} }{\gamma _{\pi } ^{3} } \left(\frac{B}{10^{-5} {\rm G}} \right)\left(\frac{T_{bh} }{{\rm GeV}} \right)^{-1} \qquad {\rm  for\; }\pi ^{\pm } 
\label{64}
\end{equation}
\begin{equation}
\frac{\Delta r_{B} }{r} \approx \frac{10^{-23} }{\gamma _{\pi } ^{3} } \left(\frac{B}{10^{-5} {\rm G}} \right)\left(\frac{T_{bh} }{{\rm GeV}} \right)^{-1} \qquad {\rm for\; }e^{\pm } . \label{65}
\end{equation}
Certainly the Galactic magnetic field, which is of order $10^{-5} {\rm \; G}$, is too weak to produce significant charge separation and hence cannot generate Langmuir oscillations. Additionally, the magnetic field strength required for significant separation would appear to be unrealistically large in any astrophysical context.

If we consider the pions, electrons and positrons to be generated in QCD jets of opening angle $\theta _{jet} $, the transverse velocity may be as large as $v_{T} \approx v\theta _{jet} \approx c\theta _{jet} $ and so the maximum separation is given by
\begin{equation} 
\frac{\Delta r_{B} }{r} \approx \frac{10^{-6} }{\gamma _{\pi } ^{3} } \left(\frac{B}{10^{-5} {\rm G}} \right)\theta _{jet} \left(\frac{r}{{\rm cm}} \right) 
\label{66} 
\end{equation} 
for $\pi ^{\pm } $ and $10^{-2} $ smaller for $e^{\pm } $. At accelerator energies, $\theta _{jet} \approx 1.29\alpha _{s} $ \cite{RW}. The maximum magnetic field-induced separation is then greater than the average separation between particles, $n^{-1/3} $, at
\begin{equation}
r \gtrsim \frac{10\gamma ^{9/4} }{\theta _{jet} ^{3/4} } \left(\frac{B}{10^{-5} {\rm G}} \right)^{-3/4} \left(\frac{T_{bh}}{{\rm GeV}} \right)^{-3/8} {\rm \; cm \qquad for\; }\pi ^{\pm }
\label{67}
\end{equation}
\begin{equation}
r \gtrsim \frac{10^{3} \gamma ^{9/4} }{\theta _{jet} ^{3/4} } \left(\frac{B}{10^{-5} {\rm G}} \right)^{-3/4} \left(\frac{T_{bh}}{{\rm GeV}} \right)^{-3/8} {\rm \; cm \qquad for\; }e^{\pm } .  \label{68}
\end{equation}
In this case, turbulence may be generated at high temperatures in a strong external magnetic field. However, only a small fraction of the emitted particles would have sufficient transverse velocity and the relevant orientation to satisfy the conditions to participate in Langmuir oscillations. We have already shown above that, once started, only the low energy $\pi ^{\pm } $ and $e^{\pm } $ can possibly sustain an MHD regime.

 We conclude that a \textit{self-induced} magnetic field and MHD regime should not arise in the charged emitted particles as they stream away from the black hole. Hence no photosphere will be generated by this mechanism. If the black hole is embedded in an \textit{external} magnetic field of sufficient strength and/or turbulence, we find that the MHD regime extends over a smaller region than Belyanin, Kocharovsky, and Kocharovsky envisage and can involve only the lowest energy emitted particles, weakening the effect. Rees has also proposed a model in which the high energy electrons emitted by an expiring black hole form an expanding conducting shell which will be braked by any ambient Galactic magnetic field \cite{R}. However, MacGibbon has found that the Rees mechanism, when re-analyzed including QCD decays, will produce a gamma-ray burst with a detection probability at Earth of much less than $1$ \cite{JHM,MC}.

\subsection{Cline and Hong Model}

D. Cline, Matthey and Otwinowski have proposed that Galactic PBHs may explain the shortest-lived gamma-ray bursts \cite{DC3}. They suggest two mechanisms for burst generation: plasma formation at $T_{bh} \approx 10{\rm \; GeV\; }-10{\rm \; TeV}$, together with the MHD instability proposed by Belyanin, Kocharovsky, and Kocharovsky \cite{BY}; or `fireball' formation at the QCD threshold $T_{bh} \sim \Lambda _{QCD} $, as proposed by Cline and Hong \cite{DH,DC1,DC2}. As we have seen, the Belyanin, Kocharovsky, and Kocharovsky mechanism is not viable in the absence of a large external magnetic field. The second mechanism postulates that the emitted particles undergo a first-order phase transition around $T_{bh} \sim \Lambda _{QCD} $, accompanied by rapid loss of the remaining black hole mass via a Hagedorn-type exponential increase in the number of hadronic degrees of freedom at $T_{bh} \sim \Lambda _{QCD} $, without the black hole temperature climbing above $\Lambda _{QCD} $. These hadrons are assumed to be directly emitted by the black hole, and not the result of decays.

The Hagedorn model was suggested by the exponential increase in hadronic resonances observed in early accelerator experiments at these energies. However, the resonances have long since been understood in terms of the more fundamental quark model, and accelerator energies well above $\Lambda _{QCD} $ have long since been achieved. The standard Hawking picture, in which the black hole emits only those particles which are fundamental on the scale of the black hole and which was discussed in detail in Section IV.K, is consistent with current accelerator observations. Thus a $T_{bh} \gtrsim \Lambda _{QCD} $ hole should directly emit quarks and gluons, which then hadronize at a distance $r\approx T_{bh} /\Lambda _{QCD} ^{2} $ from the black hole, rather than an explosive Hagedorn-type particle distribution. Similarly an $m_{\pi } \lesssim T_{bh} \lesssim \Lambda _{QCD} $ black hole should emit pions as fundamental particles. Furthermore, in the Hagedorn picture the exponential distribution of resonances are possible particle states but whether the states are occupied is determined by the available energy. As we have seen in Section IV.K, the energy per emitted quark or gluon around $T_{bh} \sim \Lambda _{QCD} $ is only $E\sim \Lambda _{QCD} $. Dense occupation of the Hagedorn resonances would require a direct coupling of the black hole mass to the Hagedorn states. If the black hole is rapidly losing mass into the Hagedorn spectrum, however, Eq.~\eqref{2} dictates that the temperature must rapidly climb above $\Lambda _{QCD} $. This would violate the condition of the original Hagedorn model \cite{RH} in which the exponential growth in the density of states results from the temperature being held constant.

Hence, based on the observed behavior in accelerators, there should be no rapid increase in the black hole evaporation rate as $T_{bh} \to \Lambda _{QCD} $ and no gamma-ray photosphere burst of the type envisioned by Cline and Hong. The `fireball' model proposed by Moss \cite{MOS}, in which a phase-transition at the $\Lambda _{QCD} $ threshold generates a confining bubble around the black hole, is also not consistent with accelerator observations. Note that the Cline and Hong `fireball' model is not related to the Heckler bremsstrahlung-induced photosphere. A Heckler-type photosphere involves only interactions of particles after emission and is not accompanied by a rapid increase in the Hawking evaporation rate.

Although our analysis rules out photospheres in such models, we cannot exclude the possibility that black hole explosions could generate short-period gamma-ray bursts by some other as yet unconsidered mechanism or by placement, for example, in unusual magnetic field configurations. The possibility that one might detect black hole explosions (and thereby test the Hawking theory) at the present epoch is so important that all such scenarios are worth exploring. In particular, it is important to examine whether short-period gamma-ray bursts have other characteristics expected of a PBH population, independent of knowledge of the specific burst production mechanism.

\section{CONCLUSIONS}

After careful examination of the Heckler photosphere model, we find that the emitted particles are not expected to interact sufficiently to form a QED or QCD photosphere around an evaporating black hole. Key to determining the fate of the emission is understanding the causality constraint and scatterings which occur within a distance $\sim 1/m_{e} $  or $\sim 1/\Lambda _{QCD} $ of the hole. Our main point is that no QED photosphere can form around a black hole because a particle must be causally connected to any particle it interacts with and requires a distance $\sim E/m_{e} ^{2} $ to form in each bremsstrahlung interaction. This means that each particle is highly unlikely to interact and generally undergoes at most one net interaction as it streams away from the hole. Neither Heckler nor Cline, Mostoslavsky, and Servant include these effects. Instead they envision a scenario in which each emitted particle undergoes a sequence of interactions. In Heckler's model these start at a radius less than $r_{brem} $ and continue to a radius greater than $r_{brem} $; in the Cline, Mostoslavsky, and Servant simulations, they start at a radius much less than $r_{brem} $ and continue to $r_{brem} $. Furthermore, the modification required to their models is not a simple one, since the causality and formation distance constraints and a corrected geometrical description must go into the analysis at the start and cannot simply be added later. Additional QCD phenomenological arguments, based on the limited energy per initial particle and availability of QCD final states, also rule out QCD photosphere (chromosphere) development around $T_{bh} \sim \Lambda _{QCD} $.

Although we find that photosphere formation is not supported, an emitted particle may occasionally undergo one net QED interaction within a distance $\sim 1/m_{e} $ of the black hole or one net QCD interaction within $\sim 1/\Lambda _{QCD} $. However, this interaction could only very slightly modify the observable $\gamma $, $p$, $\overline{p}$, $e^{+} $, $e^{-} $, $\nu $ and $\overline{\nu }$ spectra from a PBH background or an individual high-temperature black hole. The spectra remain essentially those of the standard Hawking model, with little change to the detection probability. We also find that interactions are insufficient to support other models for photosphere formation or enhanced PBH signals proposed in the literature, unless the black hole is located in an unusual ambient environment or Standard Model physics breaks down.

In this paper, we have not addressed the emission expected from higher-dimensional black 
holes which may be created at TeV or higher energies in accelerators \cite{ADMR,BF,EHM,DL,GT} or cosmic ray interactions \cite{RT,FS}. Our analysis implies that the spectra which 
were calculated, in for example Refs.~\cite{AG,CS}, assuming Heckler-type interactions between the particles emitted in the final burst of evaporation from a higher-dimensional black hole will need to be modified. In particular, the issues raised in Sections IV.B, D, F, G, H and K will need to be applied to the analysis of the spectra from higher-dimensional black holes. In the case of accelerator experiments, the relevant condition to investigate is whether ${\mathcal N} \sim 1$ is approached, because even one interaction per particle may modify the spectra significantly compared with experimental precision. A number of the issues we raise were independently recently included in the numerical simulation of the decay of CERN LHC TeV-scale black holes by Alig, Drees and Oda \cite{ADO}. They concluded that scattering between partons emitted by the higher-dimensional black hole is insufficient to form a thermalized chromosphere. Although their methodology was restricted to scattering between partons which approach each other and included a lower cutoff of $1{\rm \; GeV}$ in the exchanged transverse momentum, their results provide insight into the applicability of the Heckler scenario to TeV-scale higher-dimensional black holes. Our discussion of geometrical and causality considerations in Section IV.F and in our accompanying paper \cite{PCM} may be viewed as a partial justification of the approximation used in \cite{ADO} of only including scattering between partons which approach each other.

\begin{acknowledgments}

The authors thank Andrzej Czarnecki, Manuel Drees, Joseph Kapusta, Graham Thompson and Bryan Webber for useful discussions. JHM is grateful to Queen Mary, University of London, the Fermilab Astrophysics Group and the University of Cambridge for hospitality received at various points during this work. BJC is grateful to the Research Center for the Early Universe at the University of Tokyo for hospitality received during its completion. DNP thanks the Natural Sciences and Engineering Research Council of Canada for financial support.

\end{acknowledgments}

\end{document}